%% file: main.tex
\documentclass[conference]{IEEEtran}

\IEEEoverridecommandlockouts  

\usepackage{graphicx, import} 
\usepackage{float}
\usepackage{times}
\usepackage{multicol}
\usepackage{multirow}    
\usepackage{mathtools,amsfonts,amssymb,bm}
\usepackage{cases}
\usepackage{yfonts}
\usepackage{caption}
\usepackage{subcaption}
\usepackage[font=small]{caption}
\usepackage{colortbl}
\usepackage{xparse}
\usepackage{arydshln}
\usepackage[per-mode = symbol, binary-units=true]{siunitx}
\usepackage{soul}

\definecolor{jade}{rgb}{0.0, 0.66, 0.42}

\usepackage{tabularx,booktabs}
\newcolumntype{C}{>{\centering\arraybackslash}X} 
\makeatother
\makeatletter
\newsavebox\myboxA
\newsavebox\myboxB
\newlength\mylenA
\newcommand*\xoverline[2][0.75]{%
   \sbox{\myboxA}{$\m@th#2$}%
   \setbox\myboxB\null
   \ht\myboxB=\ht\myboxA%
   \dp\myboxB=\dp\myboxA%
   \wd\myboxB=#1\wd\myboxA
   \sbox\myboxB{$\m@th\overline{\copy\myboxB}$}
   \setlength\mylenA{\the\wd\myboxA}
   \addtolength\mylenA{-\the\wd\myboxB}%
   \ifdim\wd\myboxB<\wd\myboxA%
      \rlap{\hskip 0.5\mylenA\usebox\myboxB}{\usebox\myboxA}%
   \else
       \hskip -0.5\mylenA\rlap{\usebox\myboxA}{\hskip 0.5\mylenA\usebox\myboxB}
   \fi}
\makeatother

\usepackage[x11names]{xcolor}  

\title{Model-Based Real-Time Motion Tracking using Dynamical Inverse Kinematics on SO(3)}
\author{Lorenzo Rapetti$^{1,2}$,
Yeshasvi Tirupachuri$^{1,3}$,
Kourosh Darvish$^{1}$,
Claudia Latella$^{1}$,
Daniele Pucci$^{1}$
\thanks{$^{1}$ Dynamic Interaction Control at Istituto Italiano di
 Tecnologia, Center for Robotics Technologies, Genova,
  Italy. ({email: \tt\small name.surname@iit.it})}
\thanks{$^{2}$ Machine Learning and Optimisation, The University of Manchester,
 Manchester, United Kingdom.}
\thanks{$^{3}$ DIBRIS, University of Genova, Genova, Italy.}
}

\begin{document}

\maketitle
\thispagestyle{empty}
\pagestyle{empty}

\begin{abstract}

 This paper contributes towards the development of motion tracking algorithms for time-critical applications, proposing an infrastructure for solving dynamically the inverse kinematics of highly articulate systems such as humans. We present a method based on the integration of differential kinematics using distance measurement on SO(3) for which the convergence is proved using Lyapunov analysis. An experimental scenario, where the motion of a human subject is tracked in static and dynamic configurations, is used to validate the inverse kinematics method performance on human and humanoid models. Moreover, the method is tested on a human-humanoid retargeting scenario, verifying the usability of the computed solution for real-time robotics applications. Our approach is evaluated both in terms of accuracy and computational load, and compared to iterative optimization algorithms.

\end{abstract}


 \input{sections/introduction.tex}

\input{sections/background.tex}
 \input{sections/control_theory_based.tex}

 \input{sections/experiments.tex}
 \input{sections/results.tex}

\input{sections/conclusions.tex}



  \section*{Acknowledgments}
This work is supported by Honda R\&D Co., Ltd and by EU An.Dy Project that received funding from the European Union’s Horizon $2020$ research and innovation programme under grant agreement No. $731540$. The content of this publication is the sole responsibility of the authors. The European Commission or its services cannot be held responsible for any use that may be made of the information it contains.

\input{sections/appendix.tex}
\bibliographystyle{IEEEtran}
\bibliography{bibliography}

\end{document}

%% file: sections/introduction.tex
\section{Introduction}
\label{section:introduction}

Nowadays, real-time motion tracking has many established applications in different fields such as medicine, virtual reality, and computer gaming. Moreover, in the field of robotics there is a growing interest in human motion retargeting and imitation~\cite{Dariush2008}\cite{DeepMimic2018}. Different tracking technologies and algorithms are currently available. Among these, optical tracking techniques are more spread and have been available since the eighties~\cite{aggarwal1999human}. Inertial/magnetic tracking technologies have been available only with the advent of micromachined sensors, and ensure higher frequency of data and lower latency, that makes them suited for demanding real-time applications~\cite{zhu2004real}. The objective of motion tracking algorithm is to find the human configuration from a set of measurement. Tracking algorithms can use human body representations with different level of complexity spacing from contours~\cite{shio1991}\cite{Leung1995}, stick figure~\cite{Niyogi1994}\cite{Bharatkumar1994}, and volumes~\cite{wachter1999tracking}\cite{gall2009motion}.
For some techniques it is not required to know \textit{a priori} the shape of the model, and its identification is part of the algorithm~\cite{shio1991}\cite{Niyogi1994}.
When the human is modeled as a kinematic chain, the solution of the model inverse kinematics has a major role in the algorithm~\cite{monzani2000}\cite{pons2011outdoor}\cite{ganapathi2010real}\cite{aristidou2010motion}, hence, strategies to solve it efficiently are needed.

In the field of robotics, a common inverse kinematics problem consists in finding the mapping between the end-effector of a manipulator (task space) and the corresponding joints angle (configuration space). 
Compared to an industrial manipulator, solving the inverse kinematics for a human kinematic model can be demanding as it is a highly articulate kinematic chain. Human kinematics is redundant, it generally has a high number of degrees of freedom (DoF), and may also take into account muskoloskeletal constraints in order to ensure realistic motion. Moreover, a human moving in the space is a floating base system, and its configuration space lies on a differentiable manifold \cite{traversaro2016multibody}. 

\begin{figure}[t]
    \centering
    \includegraphics[width=0.925 \columnwidth]{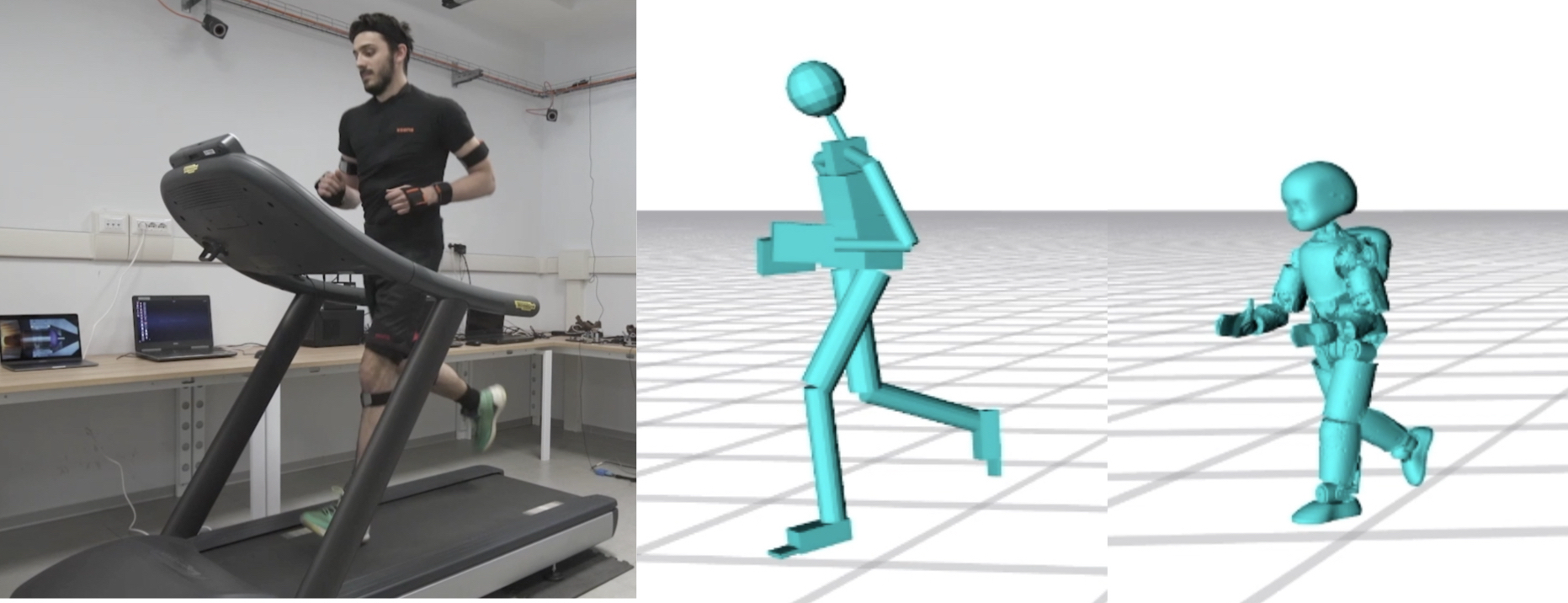}
    \caption{Motion tracking of a human running on a treadmill using an human model, on the center, and a humanoid model, on the right.}
    \label{fig:intro}
\end{figure}

Since finding an analytical closed-form solution for the inverse kinematics of a human model is not always either possible or efficient, a numerical solution is often preferred. One of the most common approaches for solving inverse kinematics is to formulate the problem as a non-linear optimization that is solved via iterative algorithms~\cite{goldenberg1985}\cite{buss2004}. This class of algorithms, referred to as \textit{instantaneous optimization} since they aim to converge to a stable solution for each time step. Although instantaneous optimization algorithms converge fast to a solution for common robotics applications, finding the solution for a human model at a sufficient rate becomes demanding for time-critical applications.
In some cases, better performances are achieved using heuristic iterative algorithms~\cite{aristidou2011fabrik}, learning algorithms~\cite{grochow2004style}, or combining analytical and numerical methods~\cite{TOLANI2000353}. An alternative to the instantaneous optimization approach consists of rephrasing the inverse kinematics problem as a control problem~\cite{sciavicco1988}. This class of algorithms is referred to as \textit{dynamical optimization} since the model configuration is controlled in order to dynamically converge over time to the sensors measurement.
From a computational point of view, the main advantage of this approach is that the solution at each time-step is computed directly with a single iteration. The absence of iterations makes the \textit{dynamical optimization} approach faster, therefore suitable for solving whole-body inverse kinematics of complex models in time-critical motion tracking applications.


This article presents a scheme for real-time motion tracking of highly articulate human, or humanoid, models. The tracking is achieved at a high frequency by using dynamic inverse kinematics optimization approach. The main contribution of this work is the application of dynamical inverse kinematics strategies to floating-base models using rotation matrix parametrization, proving the convergence of the method using Lyapunov theory, and considering model kinematic constraints by enforcing them into the differential kinematics solution. 
The implementation of the proposed scheme is tested for both human and humanoid models, during different tasks involving static and dynamic motions. The performance is compared to the results obtained using instantaneous optimization. The paper is organized as following: Section \ref{section:background} introduces the notation, human modeling, formulation of the motion tracking as inverse kinematics problem, and the dynamical optimization scheme. Section \ref{section:control_theory_based} presents our proposed dynamical inverse kinematics approach with rotation matrix parametrization and constrained inverse differential kinematics. Section \ref{section:experiments} lays the experimental details, and in Section~\ref{section:results} the results are discussed and compared to instantaneous optimization. Conclusions follow in Section \ref{section:conclusioins}.

%% file: sections/background.tex
\section{Background}
\label{section:background}

\subsection{Notation}
\begin{itemize}
    \item $\mathcal{I}$ denotes an inertial frame of reference.
    \item $I_{n \times n} \in \mathbb{R}^{n \times n}$ denotes the identity matrix of size $n$.
    \item  $\prescript{\mathcal{A}}{}{p}_\mathcal{B} \in \mathbb{R}^3$ is the the position of the origin of the frame $\mathcal{B}$ with respect to the frame $\mathcal{A}$.
    \item  $\prescript{\mathcal{A}}{}{R}_\mathcal{B} \in SO(3)$ represents the rotation matrix of the frames $\mathcal{B}$ with respect to $\mathcal{A}$.
    \item $\prescript{\mathcal{A}}{}{\omega}_\mathcal{B} \in \mathbb{R}^3$ is the angular velocity of the frame $\mathcal{B}$ with respect to $\mathcal{A}$, expressed in $\mathcal{A}$.
    \item The operator $\text{tr}(\cdot) :\mathbb{R}^{3 \times 3} \to \mathbb{R}$ denotes the \textit{trace} of a matrix, such that given $A \in \mathbb{R}^{3 \times 3}$, it is defined as $\text{tr}(A):=A_{1,1}+A_{2,2}+A_{3,3}$.
    \item The operator $\text{sk}(\cdot) :\mathbb{R}^{3 \times 3} \to so(3)$ denotes \textit{skew-symmetric} operation of a matrix, such that given $A \in \mathbb{R}^{3 \times 3}$, it is defined as $\text{sk}(A) := (A - A^\top)/2$.
    \item The operator $S(\cdot) :\mathbb{R}^{3} \to so(3)$ denotes \textit{skew-symmetric} vector operation, such that given two vectors $v,u \in \mathbb{R}^{3}$, it is defined as $v \times u = S(v)u$.
    \item The \textit{vee} operator $\cdot^{\vee} : so(3) \to \mathbb{R}^{3}$ denotes the inverse of the \textit{skew-symmetric} vector operator, such that given a matrix $A \in so(3)$ and a vector $u \in \mathbb{R}^{3}$, it is defined as $Au = A^{\vee} \times u$.
    \item The operator $\circ$ indicates the \textit{element-wise multiplication}, such that given two vectors with same length, $a=[a_1,a_2,...,a_n]$ and $b=[b_1,b_2,...,b_n]$, it is defined as $a \circ b = [a_1 b_1,a_2 b_2,...,a_n b_n]$.
    \item The operator $\text{tanh}(\cdot)$ indicates the \textit{hyperbolic tangent} function of a scalar, or the \textit{element-wise hyperbolic tangent} of a vector of scalars.
    \item The operator $\left\lVert \cdot \right\rVert_2$ indicates the L2-norm of a vector, such that given a vector $v \in \mathbb{R}^{3}$, it is defined as $\left\lVert v \right\rVert_2 = \sqrt{v_1^2+v_2^2+v_3^2}$.
\end{itemize}

\subsection{Modeling}
\label{section:background:modelling}
The human is modeled as a multi-body mechanical system composed of $n + 1$ rigid bodies, called \textit{links} that are connected by $n$ \textit{joints} with one \textit{degree of freedom} (DoF) each~\cite{latella2018towards}. Additionally, the system is assumed to be \textit{floating base}, i.e., none of the links has an \textit{a priori} constant pose with respect to the inertial frame $\mathcal{I}$. Hence, a specific frame, attached to a link of the system, is referred to as the \textit{base frame}, and denoted by $\mathcal{B}$.


The \textit{model configuration} is characterized by the position and the orientation of the \textit{base frame} along with the \textit{joint positions}. Accordingly, the configuration space lies on the Lie group $\mathbb{Q}=\mathbb{R}^3 \times SO(3) \times \mathbb{R}^n$. An element of the configuration space $q \in \mathbb{Q}$ is defined as the triplet $q = (\prescript{\mathcal{I}}{}{p}_\mathcal{B}, \prescript{\mathcal{I}}{}{R}_\mathcal{B}, s)$ where $\prescript{\mathcal{I}}{}{p}_\mathcal{B} \in \mathbb{R}^3$ and $\prescript{\mathcal{I}}{}{R}_\mathcal{B} \in SO(3)$ denote the position and the orientation of the \textit{base frame}, and $s \in \mathbb{R}^n$ denotes the joints configuration that represents the topology, i.e. shape, of the mechanical system. The position and the orientation of a frame $\mathcal{A}$ attached to the model can be obtained via geometrical forward kinematic map $h_{\mathcal{A}}(\cdot):\mathbb{Q} \to (SO(3) \times \mathbb{R}^3)$ from the \textit{model configuration}. The forward kinematics can be decomposed into position, i.e., $\prescript{\mathcal{I}}{}{{p}}_{\mathcal{A}}=h^p_{\mathcal{A}}(q)$, and orientation, i.e., $\prescript{\mathcal{I}}{}{{R}}_{\mathcal{A}} = h^o_{\mathcal{A}}(q)$, maps.

The \textit{model velocity} is characterized by the linear and the angular velocity of the \textit{base frame} along with the \textit{joint velocities}. Accordingly, the configuration velocity space lies on the group $\mathbb{V} = \mathbb{R}^3 \times \mathbb{R}^3 \times \mathbb{R}^n$. An element of the configuration velocity space $\nu \in \mathbb{V}$ is defined as $\nu = (\prescript{\mathcal{I}}{}{\mathrm{v}}_\mathcal{B}, \dot{s})$ where $\prescript{\mathcal{I}}{}{\mathrm{v}}_\mathcal{B}=(\prescript{\mathcal{I}}{}{\dot{p}}_\mathcal{B}, \prescript{\mathcal{I}}{}{\omega}_\mathcal{B}) \in \mathbb{R}^3 \times \mathbb{R}^3$ denotes the linear and angular velocity of the \textit{base frame}, and $\dot{s}$ denotes the joint velocities. The velocity of a frame $\mathcal{A}$ attached to the model is denoted by $\prescript{\mathcal{I}}{}{\mathrm{v}}_\mathcal{A}=(\prescript{\mathcal{I}}{}{\dot{p}}_\mathcal{A}, \prescript{\mathcal{I}}{}{\omega}_\mathcal{A})$ with the linear and the angular velocity components respectively. The mapping between frame velocity $\prescript{\mathcal{I}}{}{\mathrm{v}}_\mathcal{A}$ and configuration velocity $\nu$ is achieved through the \textit{Jacobian} ${J}_{\mathcal{A}}={J}_{\mathcal{A}}(q) \in \mathbb{R}^{6 \times (n+6)}$, i.e., $\prescript{\mathcal{I}}{}{\mathrm{v}}_\mathcal{A} = {J}_{\mathcal{A}}(q)  \nu$. The \textit{Jacobian} is composed of the linear part ${J}_{\mathcal{A}}^{\ell}(q)$ and the angular part ${J}_{\mathcal{A}}^{a}(q)$ that maps the linear and the angular velocities respectively, i.e., $ \prescript{\mathcal{I}}{}{\dot{p}}_\mathcal{A} = {J}_{\mathcal{A}}^{\ell}(q)  \nu$ and $\prescript{\mathcal{I}}{}{\omega}_\mathcal{A} = {J}_{\mathcal{A}}^{a}(q)  \nu$.

\subsection{Problem Statement}
Motion tracking algorithms aims to find the human configuration given a set of targets describing the kinematics of its links. The targets are the measurements of the link pose and velocity expressed in a world reference frame, and are retrieved from various sensors measurements, e.g., IMUs or image processing. The process of estimating the configuration of a mechanical system from task space measurements is generally referred to as inverse kinematics, and can be formulated as follows:


\textbf{Problem 1.} \textit{
Given a set of  $n_p$ frames $\mathcal{P} = \{ \mathcal{P}_1, \mathcal{P}_2,....\mathcal{P}_{n_p} \}$ with the associated target position $\prescript{\mathcal{I}}{}{p}_{\mathcal{P}_i}(t) \in \mathbb{R}^3$ and target linear velocity measurements $\prescript{\mathcal{I}}{}{\dot{p}}_{\mathcal{P}_i}(t) \in \mathbb{R}^3$, and given a set of $n_o$ frames $\mathcal{O} = \{ \mathcal{O}_1, \mathcal{O}_2,....\mathcal{O}_{n_o} \}$ with the associated target orientation $\prescript{\mathcal{I}}{}{{R}}_{\mathcal{O}_j}(t) \in SO(3)$ and target angular velocity measurements $\prescript{\mathcal{I}}{}{\omega}_{\mathcal{O}_j}(t) \in \mathbb{R}^3$, find the state configuration ($q(t)$,$\nu(t)$) of a model such that:
\begin{equation}\label{extended_inverse_kinematics_problem}
\begin{cases}
\prescript{\mathcal{I}}{}{{p}}_{\mathcal{P}_i}(t) = h^p_{\mathcal{P}_i}(q(t)),  & \forall i=1,\ldots,n_p \\
\prescript{\mathcal{I}}{}{{R}}_{\mathcal{O}_j}(t) = h^o_{\mathcal{O}_j}(q(t)),  & \forall j=1,\ldots,n_o \\
\prescript{\mathcal{I}}{}{\dot{{p}}}_{\mathcal{P}_i}(t) = {J}_{\mathcal{P}_i}^{\ell}(q(t))  \nu(t), & \forall i=1,\ldots,n_p \\ 
\prescript{\mathcal{I}}{}{{\omega}}_{\mathcal{O}_j}(t) = {J}_{\mathcal{O}_j}^{a}(q(t))  \nu(t), & \forall j=1,\ldots,n_o \\
A^q  s(t) \leq b^q, \\
A^{\nu}  \dot{s}(t) \leq b^{\nu},
\end{cases}
\end{equation}
where $A^q$ and $b^q$ are two constant parameters that represent the limits for the joint configuration of the model, and $A^{\nu}$ and $b^{\nu}$  constant parameters that represent the limits for the joint velocity.
}

The following quantities are defined in order to have a compact representation of \textit{Problem 1}. The targets are collected in a \textit{pose target vector} ${x}(t)$ and \textit{velocity target vector} ${\mathrm{v}}(t)$:
\begin{equation}\label{target_vectors}
{x}(t) :=
\begin{bmatrix}
\prescript{\mathcal{I}}{}{{p}}_{\mathcal{P}_1}(t) \\\
\ldots \\\
\prescript{\mathcal{I}}{}{{p}}_{\mathcal{P}_{n_p}}(t) \\
\prescript{\mathcal{I}}{}{{R}}_{\mathcal{O}_1}(t) \\
\ldots \\
\prescript{\mathcal{I}}{}{{R}}_{\mathcal{O}_{n_o}}(t)
\end{bmatrix}, \ \ 
{\mathrm{v}}(t) :=
\begin{bmatrix}
\prescript{\mathcal{I}}{}{\dot{{p}}}_{\mathcal{P}_1}(t) \\
\ldots \\
\prescript{\mathcal{I}}{}{\dot{{p}}}_{\mathcal{P}_{n_p}}(t) \\
\prescript{\mathcal{I}}{}{{\omega}}_{\mathcal{O}_1}(t) \\
\ldots \\
\prescript{\mathcal{I}}{}{{\omega}}_{\mathcal{O}_{n_o}}(t)
\end{bmatrix},
\end{equation} 
and, forward geometrical kinematics is expressed as a single vector $h(q(t))$, and Jacobians as single matrix $J(q(t))$:
\begin{equation}\label{forward_kinematics_and_jacobian}
h(q(t)) :=
\begin{bmatrix}
h^p_{\mathcal{P}_1}(q(t)) \\
\ldots \\
h^p_{\mathcal{P}_{n_p}}(q(t)) \\
h^o_{\mathcal{O}_1}(q(t)) \\
\ldots \\
h^o_{\mathcal{O}_{n_o}}(q(t))
\end{bmatrix}, \ \ 
J(q(t)) :=
\begin{bmatrix}
J^{\ell}_{\mathcal{P}_1}(q(t)) \\
\ldots \\
J^{\ell}_{\mathcal{P}_{n_p}}(q(t)) \\
J^{a}_{\mathcal{O}_1}(q(t)) \\
\ldots \\
J^{a}_{\mathcal{O}_{n_o}}(q(t))
\end{bmatrix},
\end{equation}
the set of equations in \eqref{extended_inverse_kinematics_problem} can then be reduced, using the definitions of \eqref{target_vectors} and \eqref{forward_kinematics_and_jacobian}, to the following two equations describing respectively the \textit{forward kinematics} and \textit{differential kinematics} for all the target frames:
\begin{subequations}\label{target_equations}
\begin{align}
\label{position_rotation_target_equation}
&{x}(t) = h(q(t)),  \\
\label{differential_kinematic_equation}
&{\mathrm{v}}(t) = J(q(t))  \nu(t).
\end{align}
\end{subequations}
As mentioned in Section \ref{section:introduction}, in the case of highly articulate systems, like humans, finding an analytical solution to Equations \eqref{position_rotation_target_equation} and \eqref{differential_kinematic_equation} is often vain, and a numerical optimization solution is preferred.
Hence, the optimization problem has to be properly identified, such that the minimization of the cost function implies the estimated state configuration ($\hat{q}, \hat{\nu}$) follow their ground truth.

\subsection{Dynamical Inverse Kinematics Optimization}
\begin{figure}[t]
    \centering
    \includegraphics[trim=1.5cm 0cm 0.5cm 0cm, clip=true, width=0.925 \columnwidth]{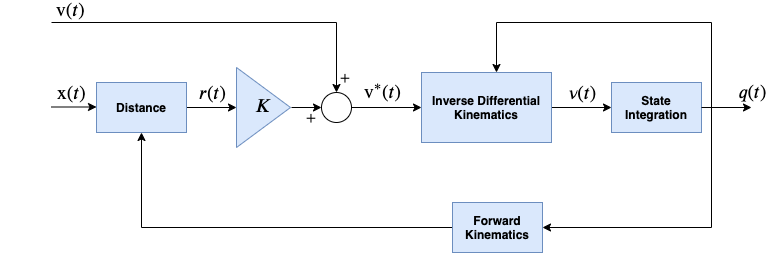}
    \caption{Dynamical optimization scheme for real-time inverse kinematics solution.}
    \label{fig:scheme}
\end{figure}

The dynamical optimization method aims to control the state configuration ($q(t),\nu(t)$) to converge to the given target measurements, i.e., to control the distance between the model and the measurements to converge to zero \cite{sciavicco1988}. The block diagram for dynamical inverse kinematics implementation is presented in Figure \ref{fig:scheme}. The scheme is composed by three main parts: a) correction of the target velocity measurements according to the residual feedback ($\mathrm{v}^*(t)=\mathrm{v}(t)+Kr(t)$), b) inversion of the model differential kinematics to obtain the state velocity $\nu(t)$, and c) integration of state velocity to obtain the configuration $q(t)$. The implementation of these three parts and the definition of the residual error are not unique and depends on different design choices. Traditionally, Euler angles or unit quaternion parametrization are used in literature for modeling floating-base systems and for defining the orientation distances. The implementation presented in the next section exploits rotation matrix parametrization for targets orientation and systems' floating-base modeling. To the best of authors' knowledge, this parametrization has not been presented in literature before.



%% file: sections/control_theory_based.tex
\section{Method}
\label{section:control_theory_based}

\subsection{Velocity Correction using Rotation Matrix}\label{velocity_correction}
The distances between the given state $q(t)$ and the pose target vector $x(t)$ are collected in a residual vector $r(q(t),x(t))$ defined as follows:
\begin{equation}\label{residual_with_distance}
r(q(t),{x}(t)) = \begin{bmatrix}\prescript{\mathcal{I}}{}{{p}}_{{\mathcal{P}_1}}(t)-h^p_{\mathcal{P}_1}(q(t)) \\\ \ldots \\\\ \prescript{\mathcal{I}}{}{{p}}_{{\mathcal{P}_{n_p}}}(t)-h^p_{\mathcal{P}_{n_p}}(q(t)) \\\ \text{sk}(h^o_{{\mathcal{O}_1}}(q(t))^T  \prescript{\mathcal{I}}{}{{R}}_{\mathcal{O}_1}(t))^{\vee} \\\ \ldots \\\ \text{sk}(h^o_{{\mathcal{O}_{n_o}}}(q(t))^T  \prescript{\mathcal{I}}{}{{R}}_{\mathcal{O}_{n_o}}(t))^{\vee} \end{bmatrix}.
\end{equation}

According to the chosen rotation matrix parametrization, the distance between two orientation measurements is computed on SO(3) with the $\text{sk}(.)^{\vee}$ operator. The distances from the target velocities $\mathrm{v}(t)$ are collected in the velocity residual vector $u(q(t), \nu(t), \mathrm{v}(t))$ defined as:  
\begin{equation}
    \label{velocity_residual_jacobian}
    u(q(t),\nu(t),{\mathrm{v}}(t)) = {\mathrm{v}}(t)-J(q(t))\nu(t).
\end{equation}

At this stage, we assume the state velocity $\nu(t)$ being the input of a dynamical system, and we want to control the system in order to drive the residual vectors towards zero.

\textbf{Lemma 1.} \textit{Assume $r(q(t),{x}(t))$ defined as in \eqref{residual_with_distance}, $u(q(t),\nu(t),{\mathrm{v}}(t))$ defined as in \eqref{velocity_residual_jacobian}, and the system
\begin{equation}\label{stable_system}
u(q(t),\nu(t),{\mathrm{v}}(t)) + K  r(q(t),{x}(t))=0,
\end{equation}
where $K \in \mathbb{R}^{(3n_p + 3n_o)} \times \mathbb{R}^{(3n_p + 3n_o)}$ is a positive definite diagonal matrix. Then, $(r,u)=(0,0)$ denotes an (almost) globally asymptotically stable equilibrium point for the system.
}

The proof is provided in appendix \textit{Lemma 1} shows that we can control the system input $\nu(t)$ so that $r(t)$ and $u(t)$ converge to zero for (almost) any initialization $q(t_0)$. Replacing the expression of $u(q(t),\nu(t))$, presented in \eqref{velocity_residual_jacobian}, in the system \eqref{stable_system}, it can be observed that the expression is linear in state velocity $\nu(t)$.
\begin{equation}\label{velocity_control_law}
    J(q(t))  \nu(t)={\mathrm{v}}(t)+K  r(q(t),{x}(t)).
\end{equation}
The rate of convergence depends on the magnitude of the elements of matrix $K$. Higher values of $K$ imply faster convergence of the system \eqref{stable_system} towards zero. However, the implementation of discrete time solution bounds the values of $K$ depending on the sampling time~\cite{ogata1995}. In fact, Equation \eqref{velocity_control_law} is solved for a discrete control input $\nu(t_k)$ from the following equation:
\begin{equation}\label{discrete_time}
    J(q(t_{k-1}))\nu(t_k)={\mathrm{v}}(t_k)+Kr(q(t_{k-1}),{x}(t_k)).
\end{equation}
Hence, according to the scheme in Figure \ref{fig:scheme}, the corrected velocity term is ${\mathrm{v}}^{*}(t_k)={\mathrm{v}}(t_k)+Kr(q(t_{k-1}),{x}(t_k))$. The way the discrete control input $\nu(t_k)$ is obtained will be discussed in Section~\ref{inverse_velocity_kinematics}.

\subsection{Constrained Inverse Differential Kinematics}\label{inverse_velocity_kinematics}
The inverse differential kinematics is the problem of inverting the differential kinematics presented in Equation \eqref{differential_kinematic_equation} in order to find the configuration state velocity $\nu(t)$ for a given set of task space velocities. In order to compute the control input $\nu(t_k)$ from \eqref{discrete_time}, it is required to solve the inverse differential kinematics for the corrected target velocity $\mathrm{v}^*(t_k)$. The solution depends on the rank of the Jacobian matrix $J(q(t_k))$, and in most of the cases it can be found only numerically as an optimization. Different strategies to solve the inverse differential kinematics can be found in literature~\cite{goldenberg1985}\cite{buss2004}\cite{sciavicco1988}\cite{Kanoun2011}, among the possible solutions, the most common approach is to use Jacobian generalized inverse~\cite{rao1972}. 
In order to take into account also the model constraints, however, a Quadratic Progamming (QP) solver is preferred since it allows to introduce a set of constraints to the problem \cite{osqp}. Hence, the inverse differential kinematics solution can be defined as the following QP optimization problem:
\begin{subequations}\label{qp_inverse_velocity_kinematics}
\begin{align}
& \underset{\nu(t_k)}{\text{minimize}} & \left\lVert\mathrm{v}^{*}(t_k) - J(q(t_{k-1}))\nu(t_k) \right\lVert_2 \\
& \text{subject to}
&  G  \nu(t_k) \leq g.
\label{eq:velocity_constraints}
\end{align}
\end{subequations}

\begin{figure}[!t]
\centering
\begin{subfigure}{0.24\textwidth}
  \includegraphics[trim=0.3cm 0cm 2.5cm 0cm, clip=true,width=0.925\columnwidth]{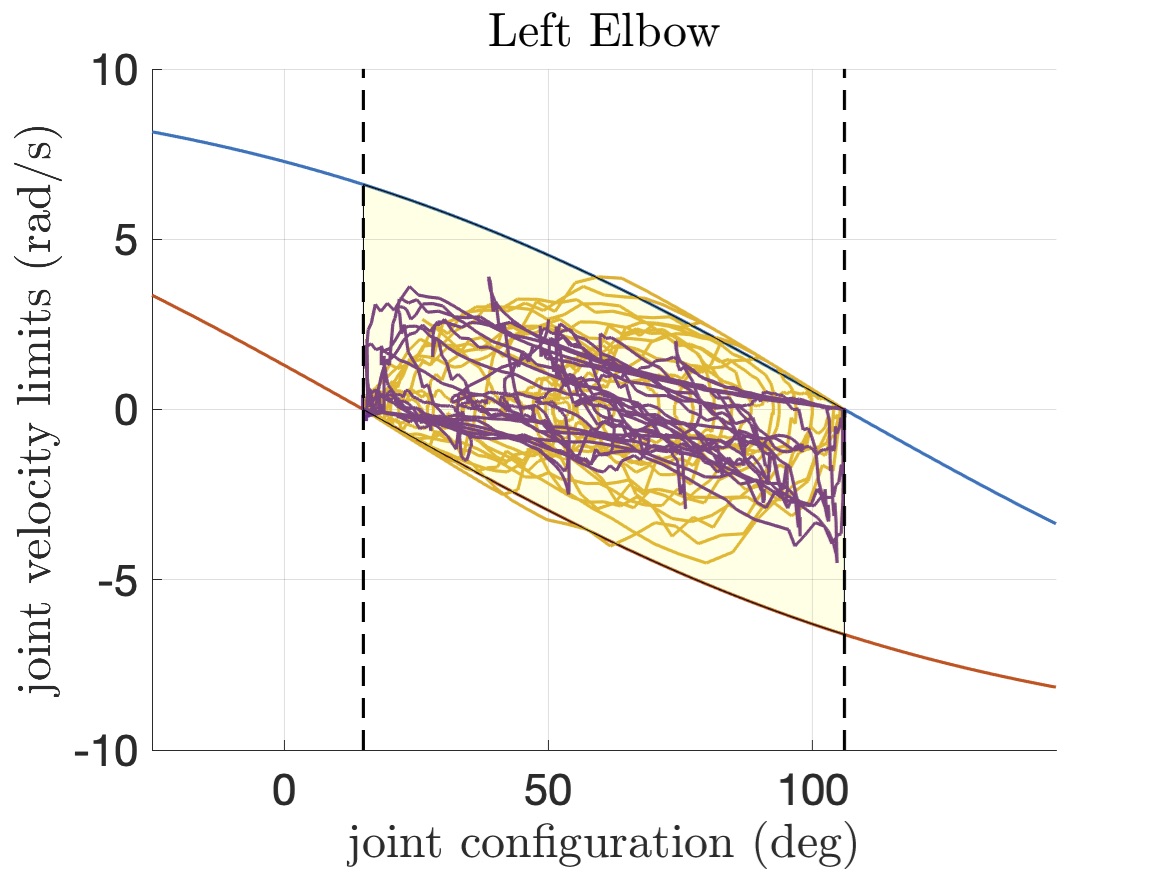}
\end{subfigure}
\begin{subfigure}{0.24\textwidth}
  \includegraphics[trim=1.2cm 0cm 2.5cm 0cm, clip=true,width=0.98\columnwidth]{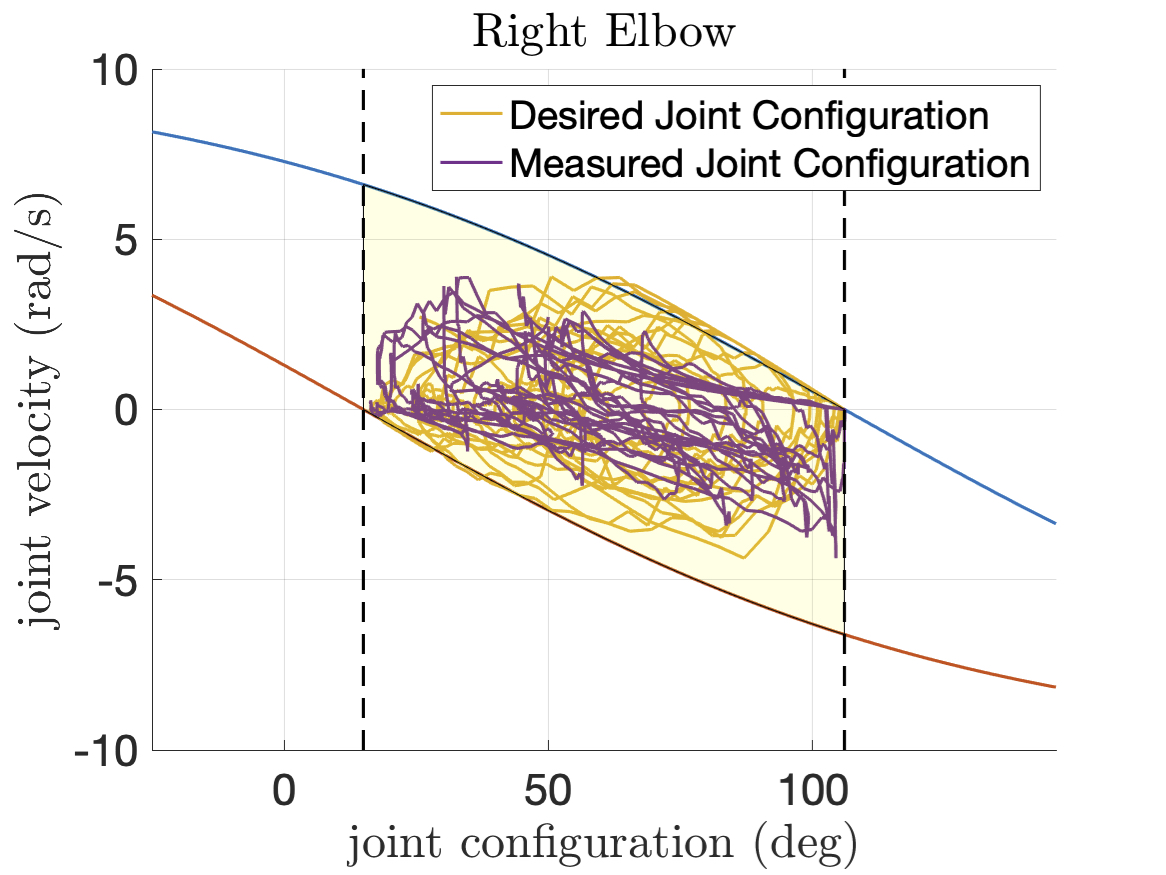}
\end{subfigure}
\caption{Constrained configuration space for the elbow joints of the iCub model. Blue and red lines represent respectively upper and lower joint velocity limits, depending on joint angle. The yellow area represents the joint configuration space. According to limit avoidance strategy, the joint velocity is bounded to be $\geq0$ when lower angle limit is reached, and $\leq0$ when upper limit is reached (angle limits are represented by dashed lines). Yellow lines represent the joint configuration trajectory computed by the inverse kinematics algorithm, while the purple line represents the trajectory tracked by a real robot.} 
\label{fig:joint_velocity_constraints}
\end{figure}

The constraints of the optimization problem defined in \eqref{eq:velocity_constraints} are linear with respect to the velocity configuration, and indeed can be used to directly enforce constraints on joint velocities $\dot{s}$.
In order to enforce system of linear constraints on joints configuration:  
\begin{equation}\label{eq:constraints}
    A q(t_k) \leq b^q,
\end{equation}
instead, the proposed joint limit avoidance strategy consists of translating \eqref{eq:constraints} to \eqref{eq:velocity_constraints} such that the dynamical constraint bounds depend on the joints configuration, i.e., $g=g(q)$. The idea is to limit the joints velocity as the joints get closer to their limits, modeling this behaviour using hyperbolic tangent function. It is assumed that for each configuration constraint, there exists a one-to-one projection to the velocity constraint space, i.e., $A^q=A^{\nu}=A$ (in case there are no constraints in a projected space, $b^{q}_i\to\infty$ and $b^{\nu}_i\to\infty$ can be approximated numerically). Therefore, the QP constraints variable used in \eqref{eq:velocity_constraints} are defined as follows:
\begin{subequations}\label{qp_constraints_equations}
\begin{align}
& G=A \\
& g=\text{tanh}(K_g(b^q-Aq))\circ b^{\nu}
\end{align}
\end{subequations}
where $K_g$ is a positive definite diagonal matrix regulating the slope of the hyperbolic tangent and $ b^{\nu}$ is a constant vector. 
Figure \ref{fig:joint_velocity_constraints} shows the effect of the proposed limit avoidance approach in constraining the joints configuration space.

\subsection{Numerical Integration}
\label{boungard_integration}
Given the configuration velocity solution $\nu(t_{k})$, it is possible to compute the state configuration $q(t_{k})$ by defining an initial configuration $q(t_0)$ and integrating over time. Base position $\prescript{\mathcal{I}}{}{p}_\mathcal{B}$ and joints configuration $s$ lie in vector space over $\mathbb{R}$ for which most of the numerical integrations methods proposed in literature can be used~\cite{davis2007}. The integration of the base angular velocity $\prescript{\mathcal{I}}{}{{\omega}}_{\mathcal{B}}(t_{k})$ is not trivial, and numerical integration errors can lead to the violation of the orthonormality condition~\cite{Gros2015} for the base orientation $\prescript{\mathcal{I}}{}{R}_\mathcal{B}$. A possible approach presented in literature is the Baumgarte stabilization~\cite{Gros2015}, the convergence of $\prescript{\mathcal{I}}{}{R}_\mathcal{B}(t_{k})$ over $SO(3)$ is ensured computing the base orientation matrix dynamics $\prescript{\mathcal{I}}{}{\dot{R}}_\mathcal{B}(t_{k}) \in \mathbb{R}^{3 \times 3}$  as follow:
\begin{subequations}\label{baumgarte_integration}
\begin{align} \label{baumgarte_correction_term}
& A(t_{k-1}) = \frac{\rho}{2} ((\prescript{\mathcal{I}}{}{R}_\mathcal{B}(t_{k-1})^{T}\prescript{\mathcal{I}}{}{R}_\mathcal{B}(t_{k-1}))^{-1}-I_{3 \times 3}), \\
\label{baumgarte_integratl_correction}
& \prescript{\mathcal{I}}{}{\dot{R}}_\mathcal{B}(t_{k}) = 
\prescript{\mathcal{I}}{}{R}_\mathcal{B}(t_{k-1}) (S(\prescript{\mathcal{I}}{}{{\omega}}_{\mathcal{B}}(t_k)) + A(t_{k-1})),
\end{align}
\end{subequations}
where $\rho \in \mathbb{R}^{+}$ is the gain regulating the convergence towards the orthonormality condition, and $\Delta t_k = t_k - t_{k-1}$ is the integration time step. The advantage of obtaining the configuration $q(t_k)$ through integration of velocity $\nu(t_k)$ is that the two estimated quantities are directly related, and continuity of the state configuration is ensured.


%% file: sections/experiments.tex
\section{Experiments}
\label{section:experiments}

\subsection{Motion Data Acquisition}
The proposed method has been implemented and tested using motion data acquired with the Xsens Awinda wearable suit~\cite{Roetenberg2009} providing pose and velocity of a 23 links human model, computed from a set of distributed Inertial Measurement Units (IMUs). The motion data is streamed through YARP middleware~\cite{yarp2006} that facilitates recording and real-time playback of data. The motion data is acquired for three scenarios with different levels of dynamicity: \textit{t-pose} where the subject stands on two feet with the arms parallel to the ground, \textit{walking} where the subject walks on a treadmill at a constant speed of $\SI{4}{\kilo\meter\per\hour}$, and \textit{running} where the subject runs on a treadmill at a constant speed of $\SI{10}{\kilo\meter\per\hour}$.

\subsection{Models}
The motion tracking is performed by using two different human models defined as in \ref{section:background:modelling}. Both the models are composed by 23 \textit{physical links} representing segments of the human body. Each physical link is attached to the next one through a certain number of rotational joint connected through \textit{dummy links}, i.e., links with dimension zero, in order to model human joints with multiple DoFs.
In one human model (\textit{Human66}) all the physical links are connected through spherical joints ($3$ rotational joints), i.e., a total of 66 DoFs and 67 links. The second model (\textit{Human48}) is based on the modeling of the human musculoskeletal system as described in clinical studies~\cite{moll1971}\cite{gerhardt1983}\cite{alison2005}, it has a reduced number of joint, i.e., 48DoFs, and takes into account human joint limits.

Additionally, we consider experiments with a model of the \textit{iCub} humanoid robot~\cite{natale2017icub}. The motivation behind this is to highlight the performance in achieving motion tracking, and motion retargeting from the human to a humanoid. The iCub model is composed of 15 physical links connected through 34 rotational joints. Model joint limits are defined according to the real robot mechanic constraints, including linear system of constraints involving multiple set of joints due to coupled joints mechanics.

The human\footnote{https://github.com/robotology/human-gazebo} and humanoid\footnote{https://github.com/robotology/icub-models} models shown in Figure \ref{fig:intro} are open-source resources. Considering that all the models have only rotational joints, the inverse kinematics problem is defined with a rotational and angular velocity target for each physical link, i.e., $n_o=23$ for the human model and $n_o=15$ for the humanoid model. Additionally, as both the models are floating base, a position and linear velocity target is used for the base frame, i.e., $n_p=1$.

\subsection{Robot Experiments}
\begin{figure*}[h]
	\centering
	\begin{subfigure}{0.24\textwidth}
		\centering
		\includegraphics[scale=0.047]{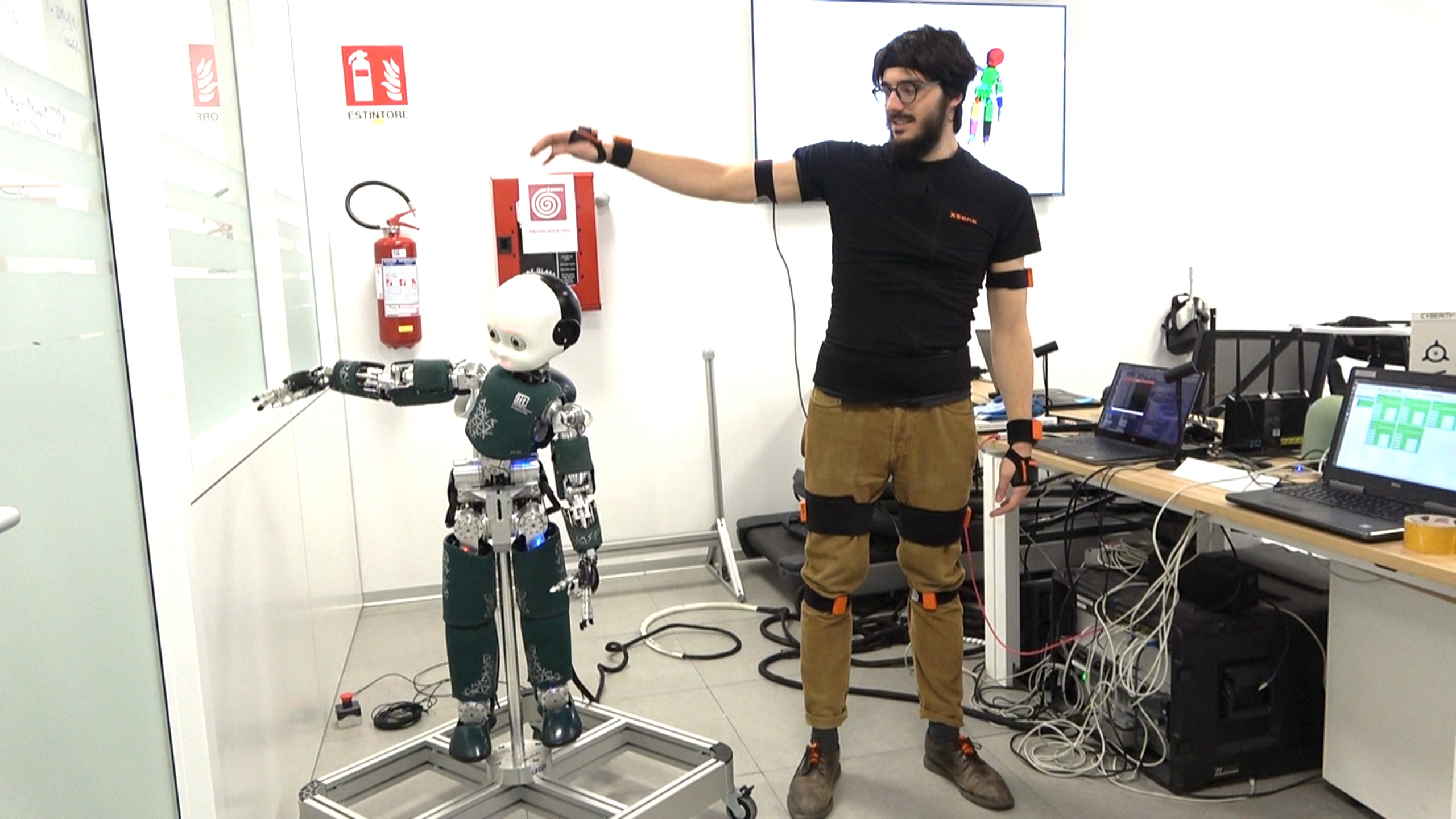}
		\label{fig:state2}
	\end{subfigure}
	\begin{subfigure}{0.24\textwidth}
		\centering
		\includegraphics[scale=0.047]{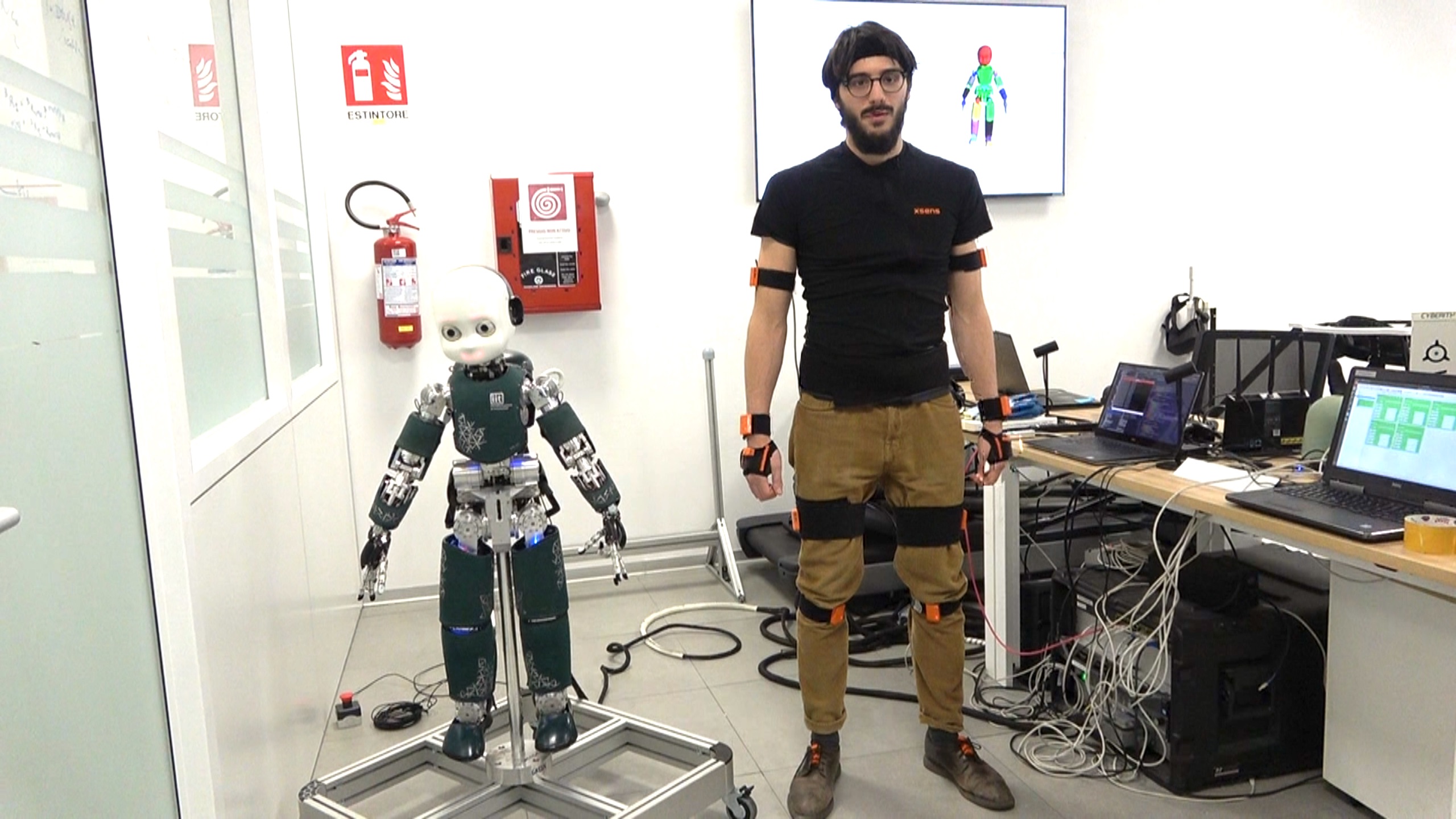}
		\label{fig:state3}
	\end{subfigure}%
	\begin{subfigure}{0.24\textwidth}
		\centering
		\includegraphics[scale=0.047]{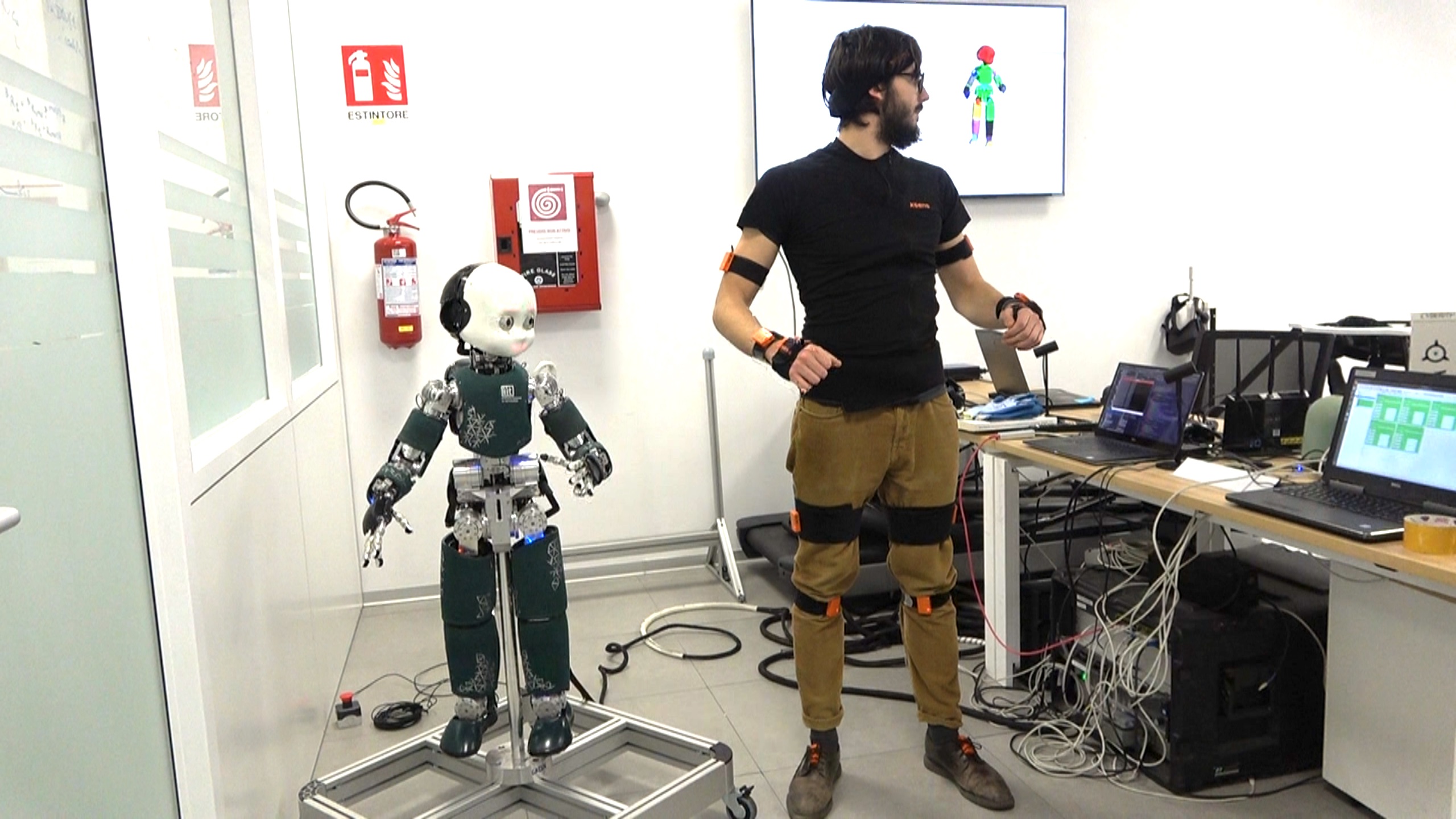}
		\label{fig:state4}
	\end{subfigure}
	\begin{subfigure}{0.24\textwidth}
		\centering
		\includegraphics[scale=0.047]{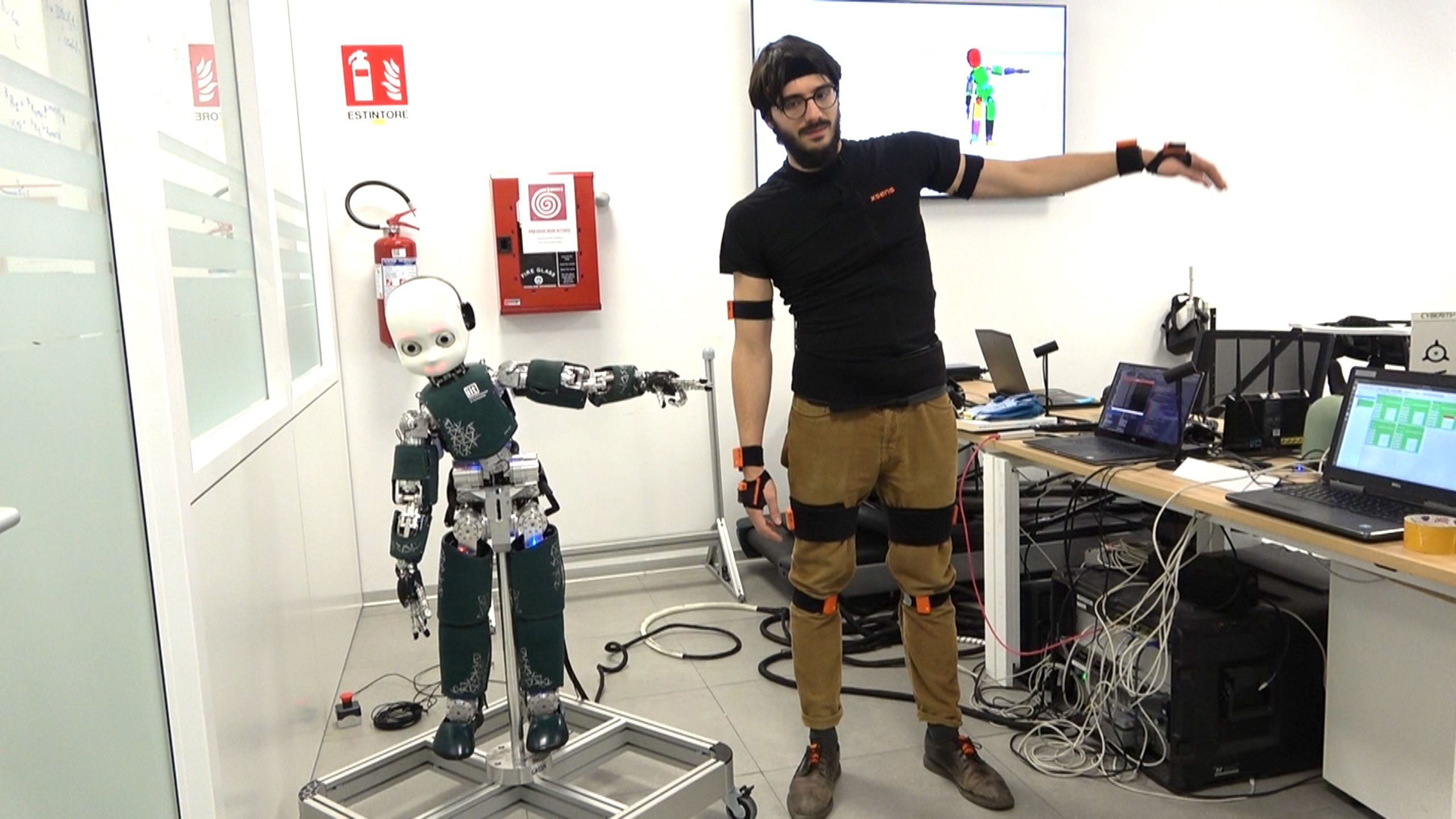}
		\label{fig:state1_human}
	\end{subfigure}
	\caption{Real-time retargeting of the human motion  to iCub humanoid robot configuration.}
	\label{fig:retargeting}
\end{figure*}

The computed joint data for the iCub model has been tested on the real robot in order to verify the feasibility of the computed robot configuration. The experimental setup involves an iCub robot fixed on a pole , as shown in Figure \ref{fig:retargeting}, controlled in position through low-level PID running at $\SI{1}{\kilo\hertz}$. The desired joint configuration is sent to the robot at $\SI{50}{\hertz}$, and it is computed real-time using the iCub model via inverse kinematics starting from human measurement data.

%% file: sections/results.tex
\section{Results}
\label{section:results}
The performance of dynamical optimization inverse kinematics solver is compared to instantaneous optimization implementations. The evaluation is done in terms of computational load and accuracy on a $\SI{2.3}{\giga\hertz}$ Intel Core i7 processor with $\SI{16}{\giga\byte}$ of RAM. The accuracy is measured with two metrics, one for the orientation targets and one for the velocity targets. The \textit{mean normalized trace error} ($MNTE$) is a dimensionless metric measuring the overall accuracy of the orientation targets:
\begin{equation}
    MNTE=\frac{1}{n_o}\sum_{j=1}^{n_o}\frac{\text{tr}(I_{3\times3}-\prescript{\mathcal{I}}{}{\hat{R}}_{\mathcal{O}_{j}}(q)^{T}\prescript{\mathcal{I}}{}{{R}}_{\mathcal{O}_{j}})}{2},
\end{equation}
where $\prescript{\mathcal{I}}{}{\hat{R}}_{\mathcal{O}_{j}}(q)$ is the estimated frame orientation given the state $q$, and the $\frac{1}{2}$ factor normalize the value of the trace between 0 and 1.
For the angular velocities, the overall error is evaluated as \textit{root mean squared error} ($RMSE$):
\begin{equation}
    RMSE=\sqrt{ \frac{1}{n_o}\sum_{j=1}^{n_o}\frac{{ \left\lVert \prescript{\mathcal{I}}{}{{\omega}}_{\mathcal{O}_{j}}- \prescript{\mathcal{I}}{}{\hat{\omega}}_{\mathcal{O}_{j}}(q,\nu) \right\rVert_2 }^2}{3}},
\end{equation}
where ${\hat{\omega}}_{\mathcal{O}_{j}}(q,\nu)$ is the estimated frame velocity given the configuration $(q,\nu)$. The computational load is evaluated as the time for computing the state $(q,\nu)$.
The statistics have been collected discarding the transient of 2 second from the initial time $t_0$.

\subsection{Instantaneous Optimization}\label{results:optimization_based}
As mentioned in Section \ref{section:introduction}, the instantaneous optimization methods solve the inverse kinematics at each time-step $t_k$ through non-linear optimization. A general formulation of the optimization problem is defined as following:
\begin{subequations}\label{nonlinear_optimization_inverse_kinematics}
\begin{align}
& \underset{q(t_k)}{\text{minimize}} & \left\lVert K_r  r(q(t_k),x(t_k))\right\lVert_2 \\
& \text{subject to} 
& A  s(t_k) \leq b
\end{align}
\end{subequations}
where $K_r$ is a weight matrix that matches the unit measurements of targets distances, and eventually assigns a weight to each of the target. A common approach for solving the non-linear optimization problem is to consider the linear approximation of the system by recalling the Jacobian matrix definition, and solve the problem iteratively~\cite{barnes1965}\cite{fletcher1968}. However, in order to enforce state configuration constraints, recent approaches make also use of convex optimization~\cite{Kanoun2011}\cite{blanchini2017convex}. As benchmark, we have implemented instantaneous inverse kinematics optimization using iDynTree~\cite{nori2015} multibody kinematics library, and the IPOPT software library for non-linear optimization~\cite{Wachter2006}. The stopping criteria for the optimization is the \textit{pose error} accuracy, and it has been tuned in order to find a solution in a time comparable to the dynamical optimization. Two different implementations have been tested. The former, referred to as \textit{whole-body optimization}, solves a single optimization problem instantiated for the whole-system. Instead, the latter instantiates the optimization process dividing the model into multiple subsystems, each consisting of exactly a pair of targets, and solves the subproblems in parallel. We refer to this implementation as \textit{pair-wise optimization}.


Looking at Figure \ref{fig:results}, it can be observed that the performance of instantaneous optimization approaches decreases as the task gets more dynamic. This is particularly evident for the computational time. In the whole-body optimization, the average computational time is higher, and it is characterized by a large variance, reaching peaks above $\SI{25}{\milli\second}$ during the running task. Concerning the pair-wised optimization, the increase of time between walking and running is less evident. However, the pair-wised optimization takes longer for finding a solution for the iCub model because of the local difference between the human and the robot kinematics.


\subsection{Dynamical Optimization}
\begin{figure}[h]
\centering
  \includegraphics[trim=2cm 0cm 2cm 1.1cm, clip=true,width=0.98\columnwidth]{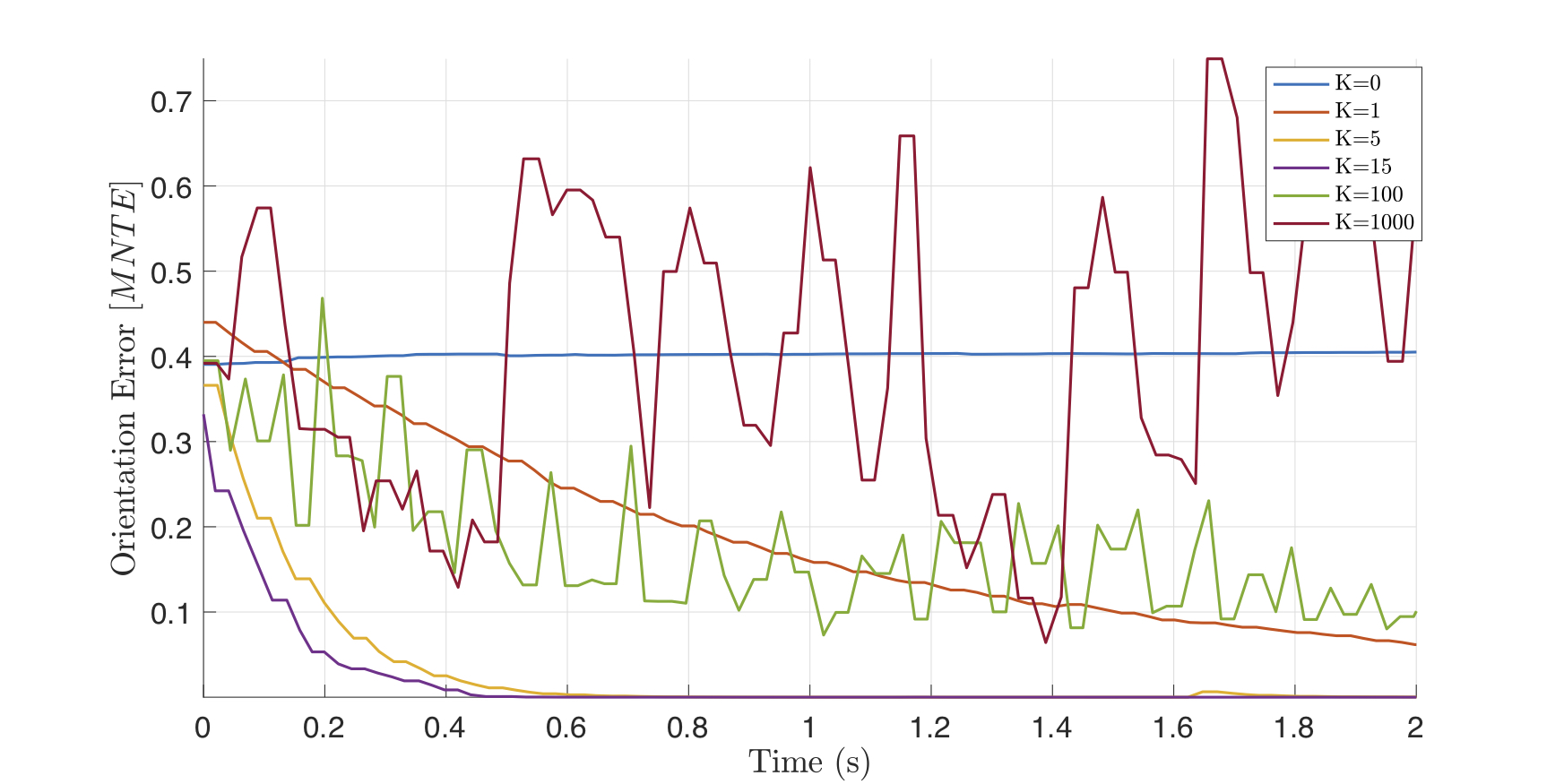}
\caption{Convergence of the dynamical inverse kinematics optimization for static \textit{T-pose} using 66 DoF model, starting from zero configuration. Convergence rate depends on the magnitude of the gain $K$.} 
\label{fig:integrationbased_convergence}
\end{figure}

The \textit{dynamical optimization} inverse kinematics has been implemented using iDynTree multibody kinematics library ~\cite{nori2015}, and the inverse differential kinematics is solved using OSQP~\cite{osqp}. Figure \ref{fig:integrationbased_convergence} highlights the rate of convergence of the error from a given initial zero configuration ($s(0)=0$) towards a target static pose. When the gain is zero, there is no velocity correction and the error remains constant. However, increasing the magnitude of $K$, the error converges to its steady state value in less then one second. A large value of the gain $K$ leads to system instability as mentioned in Section \ref{velocity_correction}.

\begin{figure*}[ht]
\centering
  \includegraphics[trim=10cm 2.7cm 7cm 2cm, clip=true, width=0.935\textwidth]{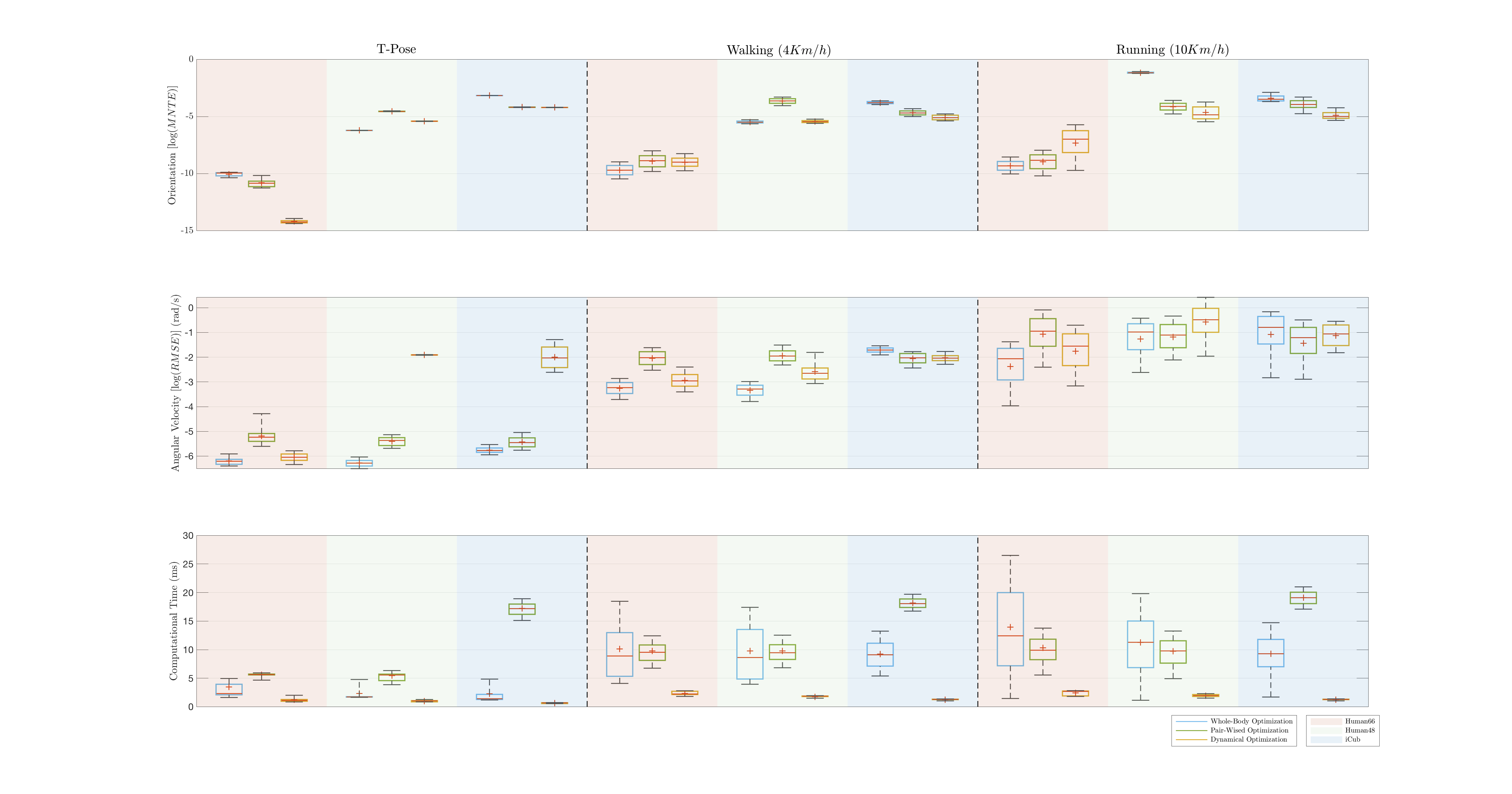}
\caption{Comparison of the performance of inverse kinematics methods (\textit{whole-body}, \textit{pair-wised}, and \textit{dynamical}) for three models (two humans, and iCub humanoid) in three different scenarios (\textit{T-pose}, \textit{Walking}, and \textit{Running}). Each line contains the boxplots for a different performance evaluation metric, on the top the overall error for the orientation targets as base 10 logarithm of \textit{mean normalized trace error}, in the middle line the overall error for the angular velocities as base 10 logarithm of \textit{root mean squared error}, and at the bottom the computational time. Logarithmic metrics allows to compare metrics characterized by different order of magnitude in the different scenarios.}
\label{fig:results}
\end{figure*}

From Figure \ref{fig:results}, the \textit{dynamical optimization} orientation error is mostly comparable with the results achieved with \textit{instantaneous optimization}. The only case in which it shows worst orientation accuracy is with the \textit{Human66} model during running task. Concerning the angular velocity error, the performances are again comparable during dynamic motion, while it is higher for the t-pose with the constrained models. This angular velocity error may be due to the fact that a corrected angular velocity is used in place of the measured link angular velocities, and the joint constraints may introduce a constant constraint error because of an unfeasible configuration. Concerning the computational load, this method seems to outperform the others not only in terms of a mean computational time, having an average always below $\SI{3}{\milli\second}$, but also for its consistency in different scenario.

\begin{figure}[h]
\centering
\begin{subfigure}{0.155\textwidth}
  \includegraphics[trim=0.2cm 0cm 1.6cm 0cm, clip=true,width=\columnwidth]{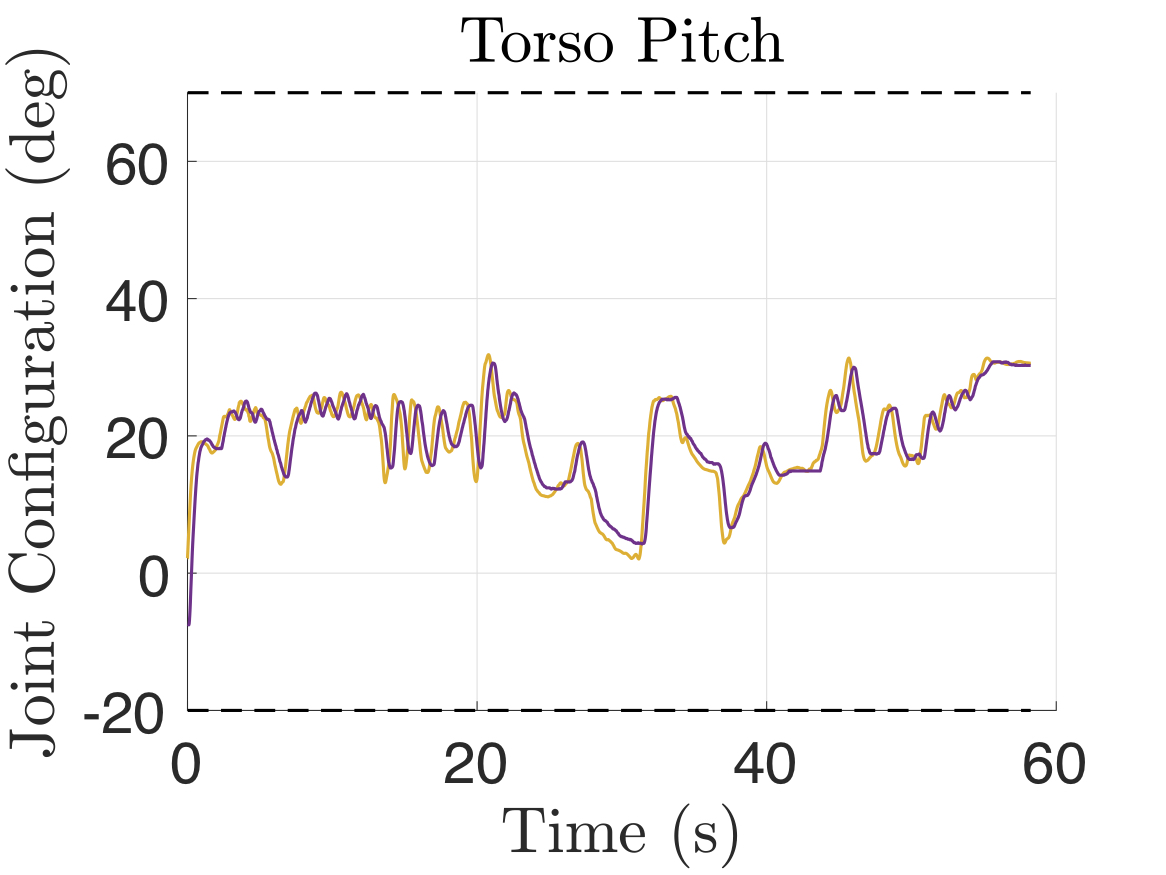}
\end{subfigure}
\begin{subfigure}{0.15\textwidth}
  \includegraphics[trim=1.2cm 0cm 1.6cm 0cm, clip=true,width=\columnwidth]{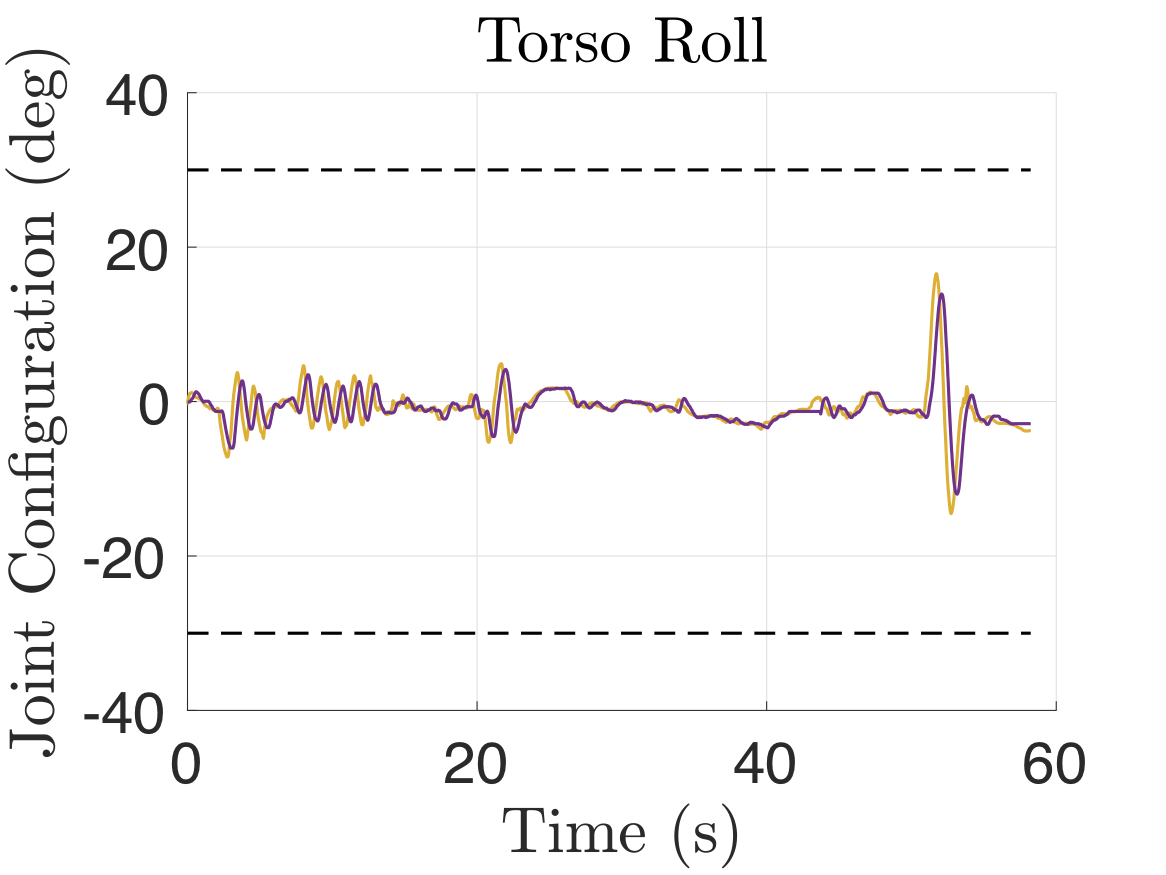}
\end{subfigure}
\begin{subfigure}{0.15\textwidth}
  \includegraphics[trim=1.2cm 0cm 1.6cm 0cm, clip=true,width=\columnwidth]{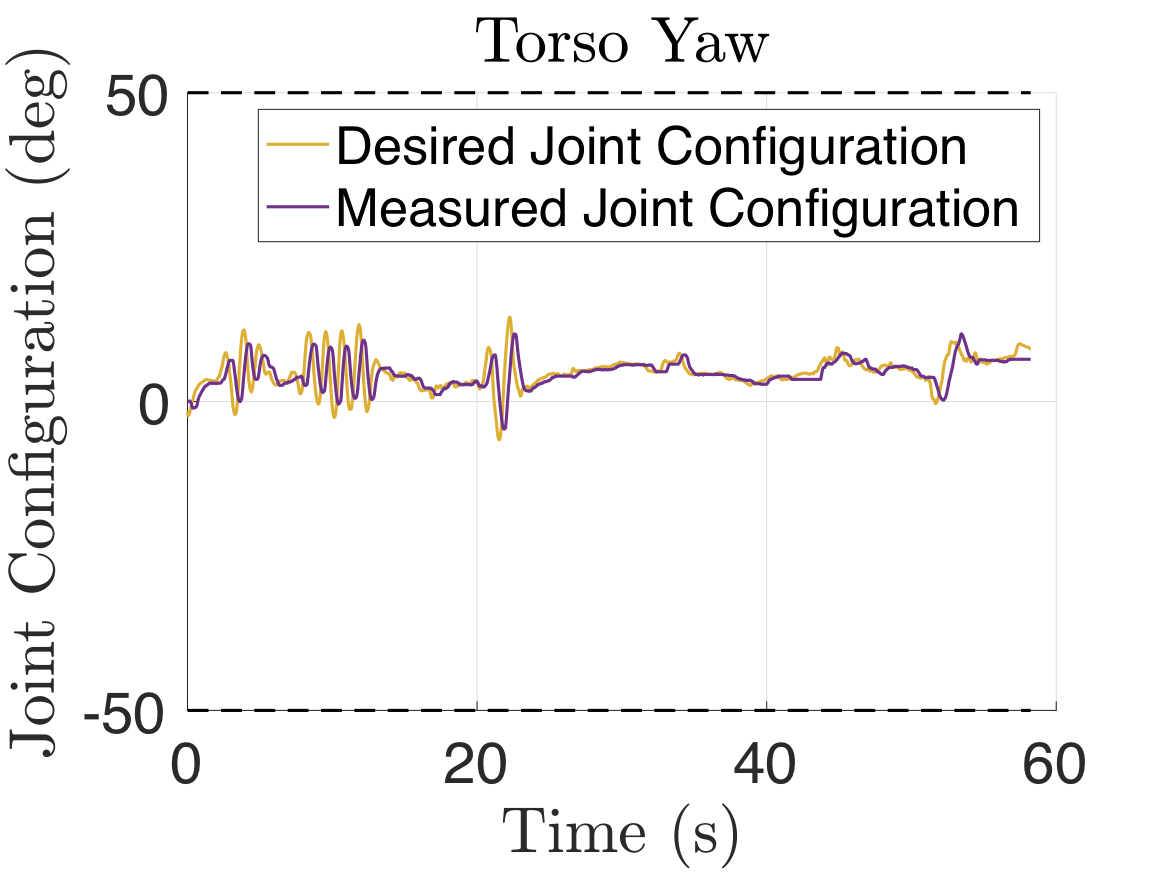}
\end{subfigure}
\begin{subfigure}{0.155\textwidth}
  \includegraphics[trim=0.2cm 0cm 1.6cm 0cm, clip=true,width=\columnwidth]{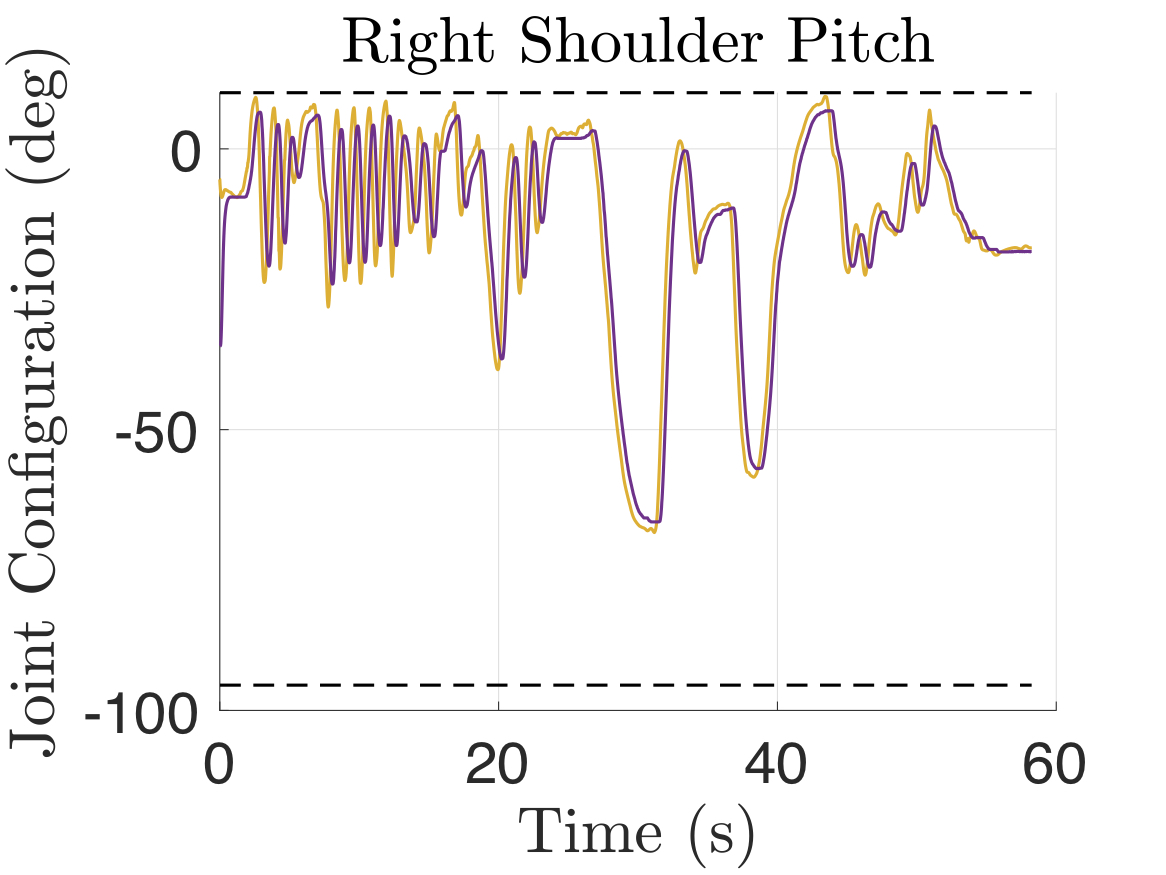}
\end{subfigure}
\begin{subfigure}{0.15\textwidth}
  \includegraphics[trim=1.2cm 0cm 1.6cm 0cm, clip=true,width=\columnwidth]{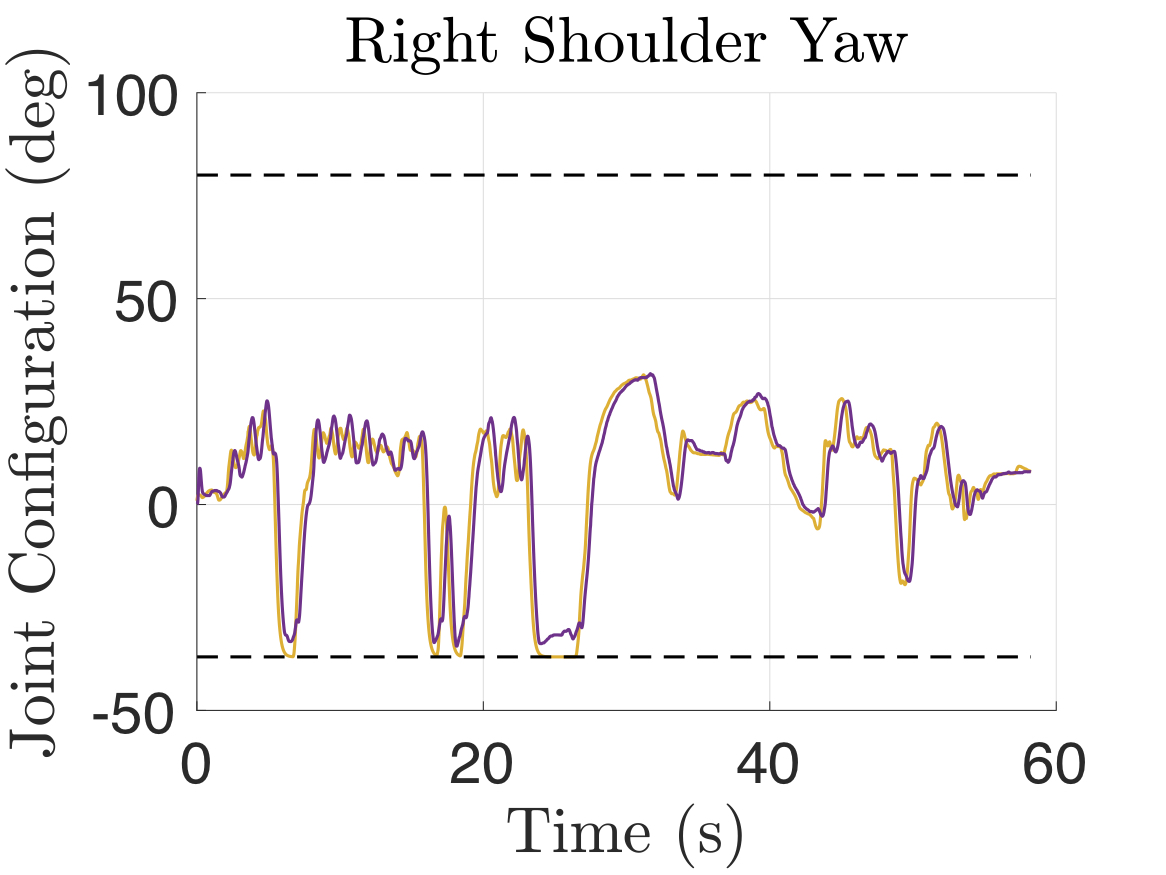}
\end{subfigure}
\begin{subfigure}{0.15\textwidth}
  \includegraphics[trim=1.2cm 0cm 1.6cm 0cm, clip=true,width=\columnwidth]{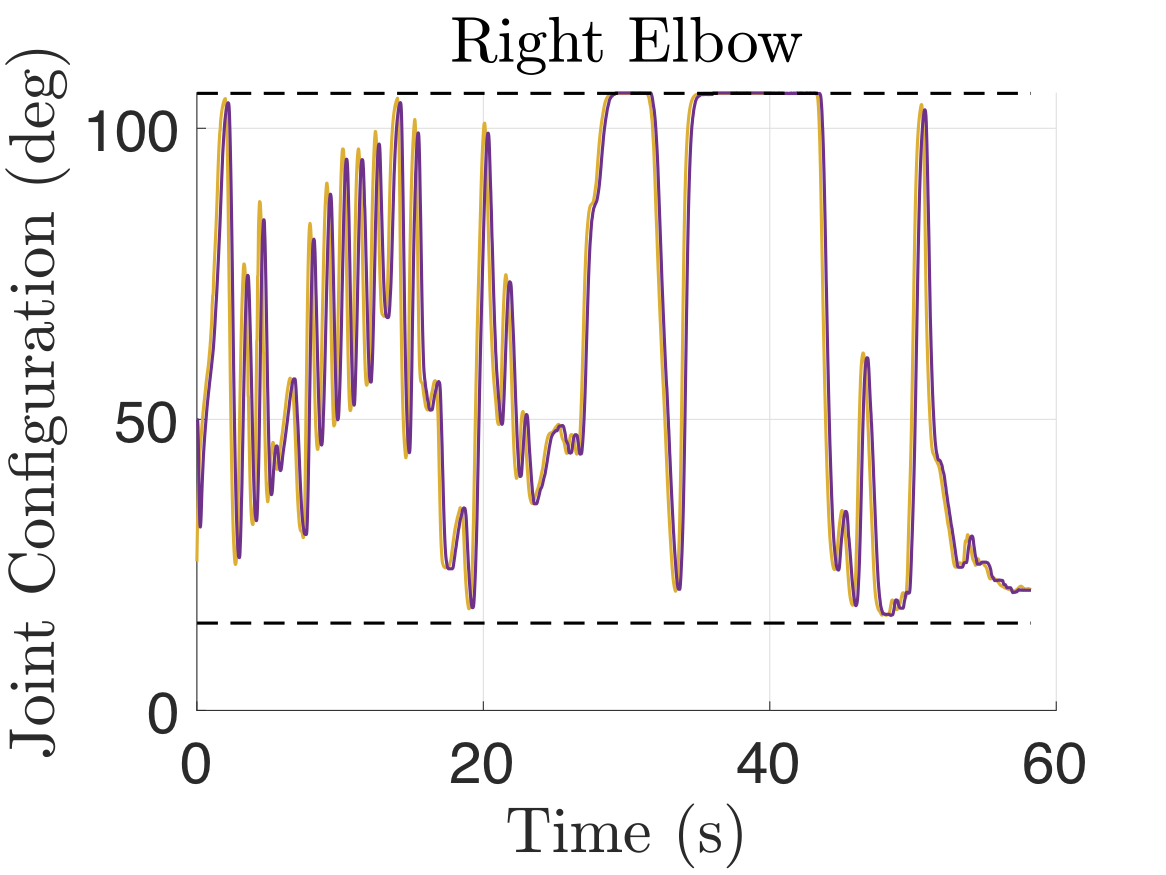}
\end{subfigure}
\caption{Joints configuration of the iCub model obtained from human motion data using dynamical inverse kinematics. The plots show both the desired joint configuration computed by inverse kinematics and sent to the robot, and the configuration measured from the robot. Dashed lines represent the joint limits.} 
\label{fig:robot-tracking}
\end{figure}

Experiments with real-robot show the feasibility of the inverse kinematics solution. The robot is able to track real-time the motion of the human by following the computed joint configuration with a delay less then $\SI{300}{\milli\second}$, as shown in Figure \ref{fig:robot-tracking} (evaluation of the controller is out of the scope of this work). The joint limit avoidance strategy successfully constraints the joint angles within the limits, and more in detail from Figure \ref{fig:joint_velocity_constraints} it can be observed how the full $(s,\dot{s})$ configuration is constrained.

%% file: sections/conclusions.tex
\section{Conclusions}
\label{section:conclusioins}
This paper presents an infrastructure for whole-body inverse kinematics of highly articulate floating-base models in real-time motion tracking applications. The theory is presented using rotation matrix parametrization of orientations, together with the proof of convergence through Lyapunov analysis. The proposed method has been implemented and the performances tested in an experimental scenario with different conditions. Differently from iterative algorithms, the \textit{dynamical optimization} requires a single iteration at each time step keeping the computational time constant, and ensures fast convergence of the error over time. Furthermore, the integration of velocities ensures obtaining a continuous and smooth solution. The method has been tested in a human-robot retargeting application to verify its usability. Its characteristics make it suitable for time-critical motion tracking applications with highly dynamic motions, where iterative algorithms may not converge in a sufficient time.

As a future work, the evaluation may be extended to a wider number of inverse kinematics algorithms, models, and experimental scenarios. Another interesting future work is to extend our method by considering the dynamics of the system.

%% file: sections/appendix.tex
\section*{Appendix: proof of Lemma 1.}
\label{section:appendix}
The system in \eqref{stable_system} can be written as follow:
\begin{equation}\label{appendix:subsystems_definition}
\begin{bmatrix}
u^{p}_{{\mathcal{P}_i}}(t) \\
\ldots \\
u^{p}_{{\mathcal{P}_{n_p}}}(t) \\
u^{o}_{{\mathcal{O}_1}}(t) \\
\ldots \\
u^{o}_{{\mathcal{O}_{n_o}}}(t) 
\end{bmatrix}
+
\begin{bmatrix}
K^{p}_1 r^{p}_{{\mathcal{P}_1}}(t) \\
\ldots \\
K^{p}_{n_p} r^{p}_{{\mathcal{P}_{n_p}}}(t) \\
K^{o}_{1} r^{o}_{{\mathcal{O}_1}}(t) \\
\ldots \\
 K^{o}_{n_o} r^{o}_{{\mathcal{O}_{n_o}}}(t) 
\end{bmatrix}
 = 0 ,
\end{equation}
where $r^{p}_{{\mathcal{P}_i}}(t)=\prescript{\mathcal{I}}{}{{p}}_{{\mathcal{P}_i}}(t)-h^p_{{\mathcal{P}_i}}(q(t))$, $r^{o}_{{\mathcal{O}_j}}(t) = \text{sk}(h^o_{\mathcal{O}_j}(q(t))^T\prescript{\mathcal{I}}{}{{R}}_{\mathcal{O}_j}(t))^{\vee}$, $u^{p}_{{\mathcal{P}_i}}(t)=\prescript{\mathcal{I}}{}{\dot{p}}_{{\mathcal{P}_i}}(t)-J^{\ell}_{{\mathcal{P}_i}}(q(t))\nu(t)$, $u^{o}_{{\mathcal{O}_j}}(t)= \prescript{\mathcal{I}}{}{{\omega}}_{{\mathcal{O}_{j}}}(t)-J^a_{{\mathcal{O}_{j}}}(q(t))\nu(t)$, $K^{p}_i$ and $K^{o}_j$ are $\mathbb{R}^{3}\times\mathbb{R}^{3}$ blocks on the diagonal of $K$.
This system can be decomposed in to a set of $n_p+n_o$ independent systems, one for each target, depending on the type of target, each subsystem is described by one of the following two equations:
\begin{subequations}\label{appendix:subsystems}
\begin{align} \label{appendix:position_system}
    &u^{p}_{{\mathcal{P}_i}}(t) + K^{p}_i r^{p}_{{\mathcal{P}_i}}(t) = 0, \\
    \label{appendix:orientation_system}
    &u^{o}_{{\mathcal{O}_j}}(t) + K^{o}_jr^{o}_{{\mathcal{O}_j}}(t) = 0.
\end{align}
\end{subequations}
The system \eqref{appendix:position_system} is a linear first order autonomous system, and for $K_i$ positive definite the equilibrium point $(r^{p}_{{\mathcal{P}_i}},u^{p}_{{\mathcal{P}_i}})=(0,0)$ is globally asymptotically stable. For the system \eqref{appendix:orientation_system} it can be proved that the equilibrium $(r^{o}_{{\mathcal{O}_j}},u^{o}_{{\mathcal{O}_j}})=(0,0)$ is an almost globally asymptotically stable equilibrium point~\cite{olfati2001}.
The almost global asymptotically stability of all the subsystems is indeed proved for the point $(r,u)=(0,0)$, thus the almost globally asymptotically stability of the equilibrium $(r,u)=(0,0)$ for the system \eqref{stable_system} is proved.

%% file: main.bbl
\begin{thebibliography}{10}
\providecommand{\url}[1]{#1}
\csname url@samestyle\endcsname
\providecommand{\newblock}{\relax}
\providecommand{\bibinfo}[2]{#2}
\providecommand{\BIBentrySTDinterwordspacing}{\spaceskip=0pt\relax}
\providecommand{\BIBentryALTinterwordstretchfactor}{4}
\providecommand{\BIBentryALTinterwordspacing}{\spaceskip=\fontdimen2\font plus
\BIBentryALTinterwordstretchfactor\fontdimen3\font minus
  \fontdimen4\font\relax}
\providecommand{\BIBforeignlanguage}[2]{{%
\expandafter\ifx\csname l@#1\endcsname\relax
\typeout{** WARNING: IEEEtran.bst: No hyphenation pattern has been}%
\typeout{** loaded for the language `#1'. Using the pattern for}%
\typeout{** the default language instead.}%
\else
\language=\csname l@#1\endcsname
\fi
#2}}
\providecommand{\BIBdecl}{\relax}
\BIBdecl

\bibitem{Dariush2008}
B.~{Dariush}, M.~{Gienger}, A.~{Arumbakkam}, C.~{Goerick}, {Youding Zhu}, and
  K.~{Fujimura}, ``Online and markerless motion retargeting with kinematic
  constraints,'' in \emph{2008 IEEE/RSJ International Conference on Intelligent
  Robots and Systems}, Sep. 2008, pp. 191--198.

\bibitem{DeepMimic2018}
\BIBentryALTinterwordspacing
X.~B. Peng, P.~Abbeel, S.~Levine, and M.~van~de Panne, ``Deepmimic:
  Example-guided deep reinforcement learning of physics-based character
  skills,'' \emph{ACM Trans. Graph.}, vol.~37, no.~4, pp. 143:1--143:14, Jul.
  2018. [Online]. Available: \url{http://doi.acm.org/10.1145/3197517.3201311}
\BIBentrySTDinterwordspacing

\bibitem{aggarwal1999human}
J.~K. Aggarwal and Q.~Cai, ``Human motion analysis: A review,'' \emph{Computer
  vision and image understanding}, vol.~73, no.~3, pp. 428--440, 1999.

\bibitem{zhu2004real}
R.~Zhu and Z.~Zhou, ``A real-time articulated human motion tracking using
  tri-axis inertial/magnetic sensors package,'' \emph{IEEE Transactions on
  Neural systems and rehabilitation engineering}, vol.~12, no.~2, pp. 295--302,
  2004.

\bibitem{shio1991}
A.~{Shio} and J.~{Sklansky}, ``Segmentation of people in motion,'' in
  \emph{Proceedings of the IEEE Workshop on Visual Motion}, Oct 1991, pp.
  325--332.

\bibitem{Leung1995}
M.~K. {Leung} and {Yee-Hong Yang}, ``First sight: A human body outline labeling
  system,'' \emph{IEEE Transactions on Pattern Analysis and Machine
  Intelligence}, vol.~17, no.~4, pp. 359--377, April 1995.

\bibitem{Niyogi1994}
{Niyogi} and {Adelson}, ``Analyzing and recognizing walking figures in xyt,''
  in \emph{1994 Proceedings of IEEE Conference on Computer Vision and Pattern
  Recognition}, June 1994, pp. 469--474.

\bibitem{Bharatkumar1994}
A.~G. {Bharatkumar}, K.~E. {Daigle}, M.~G. {Pandy}, {Qin Cai}, and J.~K.
  {Aggarwal}, ``Lower limb kinematics of human walking with the medial axis
  transformation,'' in \emph{Proceedings of 1994 IEEE Workshop on Motion of
  Non-rigid and Articulated Objects}, Nov 1994, pp. 70--76.

\bibitem{wachter1999tracking}
S.~Wachter and H.-H. Nagel, ``Tracking persons in monocular image sequences,''
  \emph{Computer Vision and Image Understanding}, vol.~74, no.~3, pp. 174--192,
  1999.

\bibitem{gall2009motion}
J.~Gall, C.~Stoll, E.~De~Aguiar, C.~Theobalt, B.~Rosenhahn, and H.-P. Seidel,
  ``Motion capture using joint skeleton tracking and surface estimation,'' in
  \emph{2009 IEEE Conference on Computer Vision and Pattern Recognition}.\hskip
  1em plus 0.5em minus 0.4em\relax IEEE, 2009, pp. 1746--1753.

\bibitem{monzani2000}
J.-S. Monzani, P.~Baerlocher, R.~Boulic, and D.~Thalmann, ``Using an
  intermediate skeleton and inverse kinematics for motion retargeting,'' in
  \emph{Computer Graphics Forum}, vol.~19, no.~3.\hskip 1em plus 0.5em minus
  0.4em\relax Wiley Online Library, 2000, pp. 11--19.

\bibitem{pons2011outdoor}
G.~Pons-Moll, A.~Baak, J.~Gall, L.~Leal-Taixe, M.~Mueller, H.-P. Seidel, and
  B.~Rosenhahn, ``Outdoor human motion capture using inverse kinematics and von
  mises-fisher sampling,'' in \emph{2011 International Conference on Computer
  Vision}.\hskip 1em plus 0.5em minus 0.4em\relax IEEE, 2011, pp. 1243--1250.

\bibitem{ganapathi2010real}
V.~Ganapathi, C.~Plagemann, D.~Koller, and S.~Thrun, ``Real time motion capture
  using a single time-of-flight camera,'' in \emph{2010 IEEE Computer Society
  Conference on Computer Vision and Pattern Recognition}.\hskip 1em plus 0.5em
  minus 0.4em\relax IEEE, 2010, pp. 755--762.

\bibitem{aristidou2010motion}
A.~Aristidou and J.~Lasenby, ``Motion capture with constrained inverse
  kinematics for real-time hand tracking,'' in \emph{2010 4th International
  Symposium on Communications, Control and Signal Processing (ISCCSP)}.\hskip
  1em plus 0.5em minus 0.4em\relax IEEE, 2010, pp. 1--5.

\bibitem{traversaro2016multibody}
S.~Traversaro and A.~Saccon, ``Multibody dynamics notation,'' \emph{Technische
  Universiteit Eindhoven, Tech. Rep}, 2016.

\bibitem{goldenberg1985}
A.~{Goldenberg}, B.~{Benhabib}, and R.~{Fenton}, ``A complete generalized
  solution to the inverse kinematics of robots,'' \emph{IEEE Journal on
  Robotics and Automation}, vol.~1, no.~1, pp. 14--20, March 1985.

\bibitem{buss2004}
S.~R. Buss, ``Introduction to inverse kinematics with jacobian transpose,
  pseudoinverse and damped least squares methods,'' \emph{IEEE Journal of
  Robotics and Automation}, vol.~17, no. 1-19, p.~16, 2004.

\bibitem{aristidou2011fabrik}
A.~Aristidou and J.~Lasenby, ``Fabrik: A fast, iterative solver for the inverse
  kinematics problem,'' \emph{Graphical Models}, vol.~73, no.~5, pp. 243--260,
  2011.

\bibitem{grochow2004style}
K.~Grochow, S.~L. Martin, A.~Hertzmann, and Z.~Popovi{\'c}, ``Style-based
  inverse kinematics,'' in \emph{ACM transactions on graphics (TOG)}, vol.~23,
  no.~3.\hskip 1em plus 0.5em minus 0.4em\relax ACM, 2004, pp. 522--531.

\bibitem{TOLANI2000353}
\BIBentryALTinterwordspacing
D.~Tolani, A.~Goswami, and N.~I. Badler, ``Real-time inverse kinematics
  techniques for anthropomorphic limbs,'' \emph{Graphical Models}, vol.~62,
  no.~5, pp. 353 -- 388, 2000. [Online]. Available:
  \url{http://www.sciencedirect.com/science/article/pii/S1524070300905289}
\BIBentrySTDinterwordspacing

\bibitem{sciavicco1988}
L.~Sciavicco and B.~Siciliano, ``A solution algorithm to the inverse kinematic
  problem for redundant manipulators,'' \emph{IEEE Journal on Robotics and
  Automation}, vol.~4, no.~4, pp. 403--410, 1988.

\bibitem{latella2018towards}
C.~Latella, M.~Lorenzini, M.~Lazzaroni, F.~Romano, S.~Traversaro, M.~A. Akhras,
  D.~Pucci, and F.~Nori, ``Towards real-time whole-body human dynamics
  estimation through probabilistic sensor fusion algorithms,'' \emph{Autonomous
  Robots}, pp. 1--13, 2018.

\bibitem{ogata1995}
K.~Ogata \emph{et~al.}, \emph{Discrete-time control systems}.\hskip 1em plus
  0.5em minus 0.4em\relax Prentice Hall Englewood Cliffs, NJ, 1995, vol.~2.

\bibitem{Kanoun2011}
O.~{Kanoun}, F.~{Lamiraux}, and P.~{Wieber}, ``Kinematic control of redundant
  manipulators: Generalizing the task-priority framework to inequality task,''
  \emph{IEEE Transactions on Robotics}, vol.~27, no.~4, pp. 785--792, Aug 2011.

\bibitem{rao1972}
C.~R. Rao, S.~K. Mitra \emph{et~al.}, ``Generalized inverse of a matrix and its
  applications,'' in \emph{Proceedings of the Sixth Berkeley Symposium on
  Mathematical Statistics and Probability, Volume 1: Theory of
  Statistics}.\hskip 1em plus 0.5em minus 0.4em\relax The Regents of the
  University of California, 1972.

\bibitem{osqp}
B.~Stellato, G.~Banjac, P.~Goulart, A.~Bemporad, and S.~Boyd, ``{OSQP}: An
  operator splitting solver for quadratic programs,'' \emph{ArXiv e-prints},
  Nov. 2017.

\bibitem{davis2007}
P.~J. Davis and P.~Rabinowitz, \emph{Methods of numerical integration}.\hskip
  1em plus 0.5em minus 0.4em\relax Courier Corporation, 2007.

\bibitem{Gros2015}
S.~{Gros}, M.~{Zanon}, and M.~{Diehl}, ``Baumgarte stabilisation over the so(3)
  rotation group for control,'' in \emph{2015 54th IEEE Conference on Decision
  and Control (CDC)}, Dec 2015, pp. 620--625.

\bibitem{Roetenberg2009}
\BIBentryALTinterwordspacing
D.~Roetenberg, H.~Luinge, and P.~Slycke, ``Xsens {MVN}: Full 6dof human motion
  tracking using miniature inertial sensors.'' [Online]. Available:
  \url{http://human.kyst.com.tw/upload/downloadfs46130703234558070.pdf}
\BIBentrySTDinterwordspacing

\bibitem{yarp2006}
G.~Metta, P.~Fitzpatrick, and L.~Natale, ``Yarp: yet another robot platform,''
  \emph{International Journal of Advanced Robotic Systems}, vol.~3, no.~1,
  p.~8, 2006.

\bibitem{moll1971}
J.~Moll and V.~Wright, ``Normal range of spinal mobility. an objective clinical
  study.'' \emph{Annals of the rheumatic diseases}, vol.~30, no.~4, p. 381,
  1971.

\bibitem{gerhardt1983}
J.~Gerhardt, ``Clinical measurements of joint motion and position in the
  neutral-zero method and sftr recording: basic principles,''
  \emph{International rehabilitation medicine}, vol.~5, no.~4, pp. 161--164,
  1983.

\bibitem{alison2005}
M.~Alison~Middleditch, M.~Jean~Oliver \emph{et~al.}, \emph{Functional anatomy
  of the spine}.\hskip 1em plus 0.5em minus 0.4em\relax Elsevier Health
  Sciences, 2005.

\bibitem{natale2017icub}
L.~Natale, C.~Bartolozzi, D.~Pucci, A.~Wykowska, and G.~Metta, ``icub: The
  not-yet-finished story of building a robot child,'' \emph{Science Robotics},
  vol.~2, no.~13, p. eaaq1026, 2017.

\bibitem{barnes1965}
J.~Barnes, ``An algorithm for solving non-linear equations based on the secant
  method,'' \emph{The Computer Journal}, vol.~8, no.~1, pp. 66--72, 1965.

\bibitem{fletcher1968}
R.~Fletcher, ``{Generalized Inverse Methods for the Best Least Squares Solution
  of Systems of Non-Linear Equations},'' \emph{The Computer Journal}, vol.~10,
  no.~4, pp. 392--399, 02 1968.

\bibitem{blanchini2017convex}
F.~Blanchini, G.~Fenu, G.~Giordano, and F.~A. Pellegrino, ``A convex
  programming approach to the inverse kinematics problem for manipulators under
  constraints,'' \emph{European Journal of Control}, vol.~33, pp. 11--23, 2017.

\bibitem{nori2015}
F.~Nori, S.~Traversaro, J.~Eljaik, F.~Romano, A.~Del~Prete, and D.~Pucci,
  ``icub whole-body control through force regulation on rigid noncoplanar
  contacts,'' \emph{Frontiers in Robotics and AI}, vol.~2, no.~6, 2015.

\bibitem{Wachter2006}
A.~W{\"a}chter and L.~T. Biegler, ``On the implementation of an interior-point
  filter line-search algorithm for large-scale nonlinear programming,''
  \emph{Mathematical Programming}, vol. 106, no.~1, pp. 25--57, Mar 2006.

\bibitem{olfati2001}
R.~Olfati-Saber, ``Nonlinear control of underactuated mechanical systems with
  application to robotics and aerospace vehicles,'' Ph.D. dissertation,
  Massachusetts Institute of Technology, 2001.

\end{thebibliography}
